\documentclass[paper]{JHEP3}
\usepackage{epsfig}
\usepackage{amsmath}
\usepackage{amssymb}
\usepackage{amsbsy}
\usepackage{enumerate}
\usepackage{url}

\newcommand\sss{\mathchoice%
{\displaystyle}%
{\scriptstyle}%
{\scriptscriptstyle}%
{\scriptscriptstyle}%
}

\def\beq{\begin{equation}}
\def\beqn{\begin{eqnarray}}
\def\eeq{\end{equation}}
\def\eeqn{\end{eqnarray}}

\newenvironment{enumerateroman}{\begin{enumerate}[i.] }{\end{enumerate}}

\newcommand\HERWIG{{\tt HERWIG}}

\newcommand\PYTHIA{{\tt PYTHIA}}

\newcommand\MCFM{{\tt MCFM}}

\def\ord#1{{\cal O}\(#1\)}

\newcommand\pt{p_{\sss\rm T}}
\newcommand\kt{k_{\sss\rm T}}
\newcommand\etmiss{{E_{\sss\rm T}\!\!\!\!\!\!\!/\,\,\,\,}}
\newcommand\TeV{\rm TeV}
\newcommand\GeV{{\rm GeV}}
\newcommand\pb{{\rm pb}}

\newcommand\ptmin{{\pt^{\min}}}

\def\lq{\left[} 
\def\rq{\right]} 
\def\rg{\right\}} 
\def\lg{\left\{} 
\def\({\left(} 
\def\){\right)} 
\newcommand\as{\alpha_{\sss\rm S}}

\newcommand\stepf{\theta}
\newcommand\POWHEG{{\tt POWHEG}}
\newcommand\POWHEGBOX{{\tt POWHEG BOX}}

\newcommand\MCatNLO{{\tt MC@NLO}}

\newdimen\hbigcirc
\newdimen\wbigcirc

\newcommand\Kinnpo{{\bf \Phi}_{n+1}}
\newcommand\Kinn{{\bf \Phi}_n}

\newcommand\Rad{\Phi_{\rm rad}}

\newcommand\CA{C_{\sss\rm A}}

\newcommand\CF{C_{\sss\rm F}}

\newcommand\Wbb{Wb\bar{b}}
\newcommand\Wmbbdec{W^-b\bar{b}\to l^-\bar{\nu}\,b\bar{b} }

\newlength{\wfig}
\newlength{\hfig}
\newcount\minutes 
\newcount\scratch 
 
\def\timestamp{%
\scratch=\time 
\divide\scratch by 60 
\edef\hours{\the\scratch} 
\multiply\scratch by 60 
\minutes=\time 
\advance\minutes by -\scratch 
---$\,$\hours:\null 
\ifnum\minutes< 10 0\fi 
\the\minutes} 
 
\preprint{}

\title{$\boldsymbol{W^\pm b\bar{b}}$ production in \textsc{POWHEG}}
  
\author{Carlo Oleari\\
  Universit\`a di Milano-Bicocca and INFN, Sezione di Milano-Bicocca\\
  Piazza della Scienza 3, 20126 Milan, Italy\\
  E-mail: \email{Carlo.Oleari@mib.infn.it}}
\author{Laura Reina\\
  Department of Physics, Florida State University\\
  Tallahassee, FL 32306-4350, U.S.A.\\
  E-mail: \email{reina@hep.fsu.edu}}
\abstract{
We present an implementation of the next-to-leading order hadronic production
of a $W^\pm$ boson in association with a pair of massive bottom quarks in the
framework of \POWHEG{}, a method to consistently interface NLO QCD
calculations with shower Monte Carlo generators.  The process has been
implemented using the \POWHEGBOX{}, an automated computer code that
systematically applies the \POWHEG{} method to NLO QCD calculations. Spin
correlations in the decay of the $W^\pm$ boson into leptons have been taken
into account using standard approximated techniques. We present
phenomenological results for $W b\bar{b}\to l\nu b\bar{b}$ production, at
both the Tevatron and the LHC, obtained by showering the \POWHEG{} results
with \PYTHIA{} and \HERWIG{}, and we discuss the outputs of the two different
shower Monte Carlo programs.
}

\keywords{QCD, Monte Carlo, NLO Computations, Resummation, Collider Physics\\
\vspace{2cm} 
\today \timestamp \hfill 
}

\begin{document}

\section{Introduction}

The production of a $W^\pm$ boson in association with a pair of massive
bottom quarks ($b$ and $\bar{b}$), contributing to both the $W+1\,b$-jet and
$W+2\,b$-jets signatures, represents both an interesting Standard Model
signal and one of the most important background processes for single-top
production and Higgs searches.

The cross sections for $W$ boson production with bottom quarks has been
measured at the Tevatron $p \bar{p}$ collider at Fermilab by both the
CDF~\cite{Aaltonen:2009qi} and D0~\cite{Abazov:2004jy} Collaborations.
As more data will be collected and analyzed by the Tevatron Collaborations,
we will gain increasing precision in the cross section measurements and we
will have a unique opportunity to test and improve the theoretical
description of heavy-quark jets at hadron colliders by performing a thorough
comparison between the Tevatron experimental data and existing theoretical
predictions.
Studying the same cross sections in the very different kinematic regimes
available at the LHC $pp$ collider will then be of great interest and will
represent a crucial test of our understanding of QCD at high-energy
colliders.

Moreover, the production of a $W$ boson with one or two $b$ jets represents a
crucial irreducible background for both single-top production
($p\bar{p},pp\rightarrow t\bar{b},\bar{t}b$) and Higgs-boson associated
production ($WH$), followed by the decay $H\rightarrow b\bar{b}$. We remind
that $WH$ associated production is the most sensitive Higgs-boson production
channel at the Tevatron, while it is a difficult, but very important, channel
at the LHC, where, in particular kinematic regions (boosted $H$), can provide
essential additional signal for the detection of a low-mass Higgs
boson~\cite{PiacquadioThesis,Butterworth:2008iy,ATLAS-2009-068}. 


It is therefore crucial to have the $W+b$-jets background theoretically under
good control.  Several steps have already been taken towards this goal. At
the parton level, the production of a $W$ boson with up to two jets, one of
which is a $b$ jet, has been calculated including next-to-leading-order~(NLO)
QCD corrections in the variable-flavor scheme~\cite{Campbell:2006cu}, while
the production of a $W$ boson with two $b$ jets has been computed at NLO in
QCD using the fixed-flavor scheme, first in the massless $b$-quark
approximation~\cite{Bern:1997sc, Bern:1996ka, Ellis:1998fv, Campbell:2002tg}
and more recently including full $b$-quark mass
effects~\cite{FebresCordero:2006sj, Cordero:2008ce, Cordero:2009kv,
  Badger:2010mg}.  The two calculations have been combined in
ref.~\cite{Campbell:2008hh} to provide the most accurate theoretical
predictions for $W+1\,b$-jet production. The comparison with the experimental
measurement of the total cross section for $W$ plus at least one $b$ jet
shows a clear discrepancy~\cite{Aaltonen:2009qi, Cordero:2010de,
  Cordero:2010qn} which should be further investigated, given the importance
of this background, in particular for Higgs boson searches.
A few predictions for $W+2 b$ jets at NLO in QCD can be found in
ref.~\cite{FebresCordero:2006sj} for the Tevatron and in
ref.~\cite{Cordero:2009kv} for the LHC, and are available to the experimental
community for comparison.


The Standard Model prediction for $Wb\bar{b}$ production could be further
improved by properly interfacing the parton level NLO calculation with a
parton shower~(PS) simulation. An event generator of this nature should also
be beneficial in understanding other experimental systematics for which
parton shower simulations are relied upon, including the transverse momentum
and rapidity distributions of $b$ jets and hard non-$b$ jets, as well as the
angular separation and invariant-mass distribution of the two $b$-jet system,
often used in experimental analyses to enhance the signal to background
ratio.

In recent times, the construction of these NLO+PS event generators has become
viable: see, e.g., \MCatNLO{}~\cite{Frixione:2002ik} and
\POWHEG{}~\cite{Nason:2004rx,Frixione:2007vw}. The effectiveness of the
\POWHEG{} approach has been demonstrated successfully and studied in some
detail through its application to a substantial array of hadron collider
processes (see, for example,
\cite{Alioli:2008gx, 
Alioli:2008tz, 
Nason:2009ai,  
Alioli:2009je, 
Hoche:2010pf,   
Re:2010bp,     
Alioli:2010qp, 
Alioli:2010xa, 
Kardos:2011qa, 
Melia:2011gk,  
Hoche:2010kg,
Hamilton:2009za,   
Hamilton:2010mb,   
D'Errico:2011um,
D'Errico:2011sd}). 
In this paper we interface the NLO calculation of the cross section for
$\Wbb$ production to a shower Monte Carlo program within the \POWHEG{}
framework.  This is the first time that such calculation has been performed:
more specifically, we have implemented this process using the
\POWHEGBOX{}~\cite{Alioli:2010xd}, a general computer code framework for
embedding NLO calculations into shower Monte Carlo programs according to the
\POWHEG{} method.  Spin correlations in the decay of the vector boson into
leptons have been taken into account using standard approximated
techniques~\cite{Frixione:2007zp}. We have checked that we have very good
agreement between our approximated result and the NLO calculation for $\Wbb
\to l\nu b\bar{b}$ production of ref.~\cite{Badger:2010mg}, where spin
correlations have been taken into account exactly. 

The paper is organized as follows. In section~\ref{sec:POWHEG_intro} we give
a brief description of the \POWHEG{} implementation of the $\Wbb$ process and
of the needed ingredients.  Since we use the \POWHEGBOX{} to implement our
process, we refer the reader to the \POWHEGBOX{}
publication~\cite{Alioli:2010xd}, and we report here only those aspects of
the implementation that are particularly relevant to the process in question.
In section~\ref{sec:theoretical_intro}, after briefly recalling how \POWHEG{}
generates an event and how the \POWHEGBOX{} deals with the presence of
Born-zero configurations, we illustrate how we have implemented the decay of
the $W$ boson. In section~\ref{sec:phenomenology} we present and discuss a
few results for the Tevatron and the LHC, where the \POWHEG-generated events
are showered by \PYTHIA{} and \HERWIG{}. Finally, in
section~\ref{sec:conclusions}, we summarize our results and give our final
remarks.

\section{Construction of the \POWHEG{} implementation}
\label{sec:POWHEG_intro}

\subsection{The next-to-leading order cross sections}
\label{sec:nlo_calculation}

\begin{figure}[htb]
\begin{center}
\includegraphics[scale=0.75]{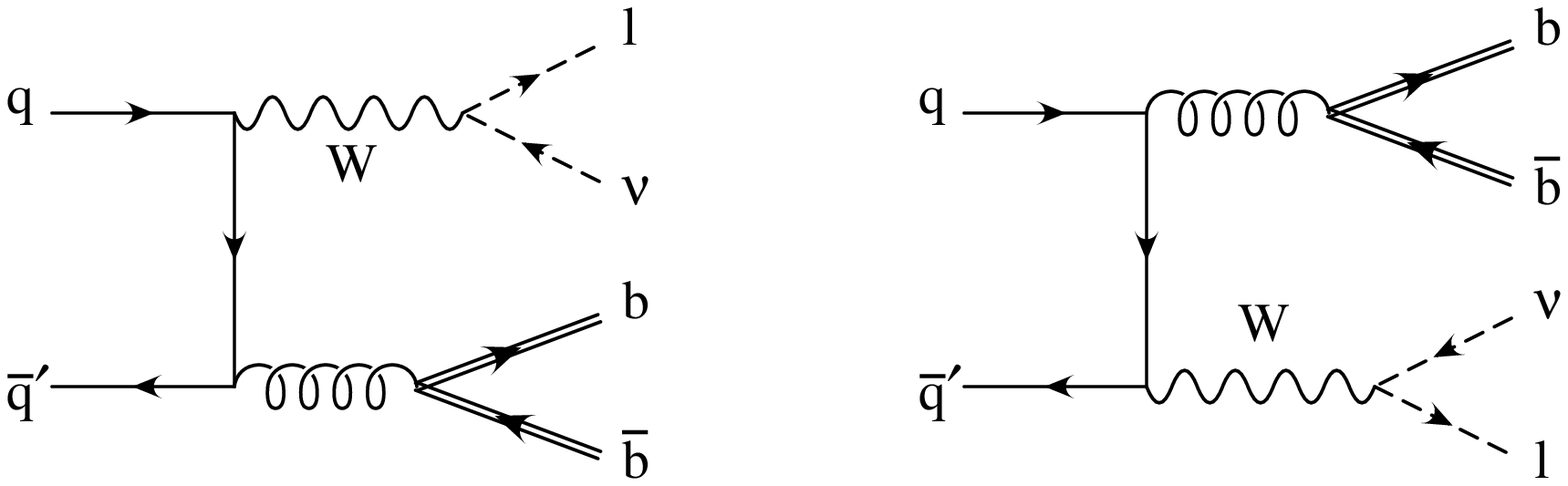} 
\caption{The tree-level Feynman diagrams for $q\bar{q}^\prime\rightarrow
  Wb\bar{b}$.} 
\label{fig:tree_level}
\end{center}
\end{figure}

The NLO QCD corrections to $q\bar{q}^\prime\rightarrow Wb\bar{b}$ production
consist of both one-loop virtual corrections to the tree level processes
depicted in fig.~\ref{fig:tree_level} and one-parton real radiation from both
the initial- and final-state quarks, i.e.~$q\bar{q}^\prime\rightarrow
Wb\bar{b}+g$. At the same order, the $qg(g \bar{q}^\prime)\rightarrow
Wb\bar{b}+q^\prime(\bar{q})$ processes also need to be included.
The $\ord{\alpha_s^2}$ virtual corrections consist of self-energy, vertex,
box and pentagon diagrams. Dimensional regularization is used to regularize
both UV and IR singularities. The UV singularities are cancelled by
introducing a suitable series of counterterms. We renormalize the wave
functions of the external massive quarks in the on-shell scheme, and the
strong coupling constant $\alpha_s$ in the mixed scheme of
ref.~\cite{Collins:1978wz}, in which the heavy-flavour top loop is subtracted
at zero momentum, while all the other flavours are subtracted in the
$\overline{\rm MS}$ scheme.  Self-energy, vertex, box and pentagon diagrams
contain IR divergences that combine and cancel against the soft and collinear
divergences in the real-emission corrections, when computing infrared-safe
quantities with renormalized parton distribution functions~(pdf).  More
details on the calculation of the $\ord{\alpha_s^2}$ virtual corrections can
be found in refs.~\cite{FebresCordero:2006sj, Cordero:2008ce,
  Cordero:2009kv}.

While in refs.~\cite{FebresCordero:2006sj, Cordero:2008ce, Cordero:2009kv}
the $W$ boson is not decayed and is produced on-shell, the \POWHEGBOX{}
implementation of $Wb\bar{b}$ production includes the leptonic decay of the
$W$, in an approximated way, as described in section~\ref{sec:w_decay}. To
implement this, the analytic results of refs.~\cite{FebresCordero:2006sj,
  Cordero:2008ce, Cordero:2009kv} have been modified to lift the $W$ on-shell
condition and the invariant mass of the $W$ boson has been generated
according to a Breit-Wigner distribution function.

\subsection{The \POWHEGBOX{} ingredients}
\label{sec:powheg_implementation}

We have implemented the NLO QCD cross section for $Wb\bar{b}$ production into
the \POWHEGBOX{} environment by providing the following ingredients:

\begin{enumerateroman}
\item the list of all flavour structures of the Born processes;

\item the list of all flavour structures of the real processes;

\item \label{item:born} the squared Born amplitude ${\cal B}$ and the colour
  correlated ones ${\cal B}_{ij}$\footnote{We notice that the $B^{\mu\nu}$
    spin-correlated Born amplitudes are zero, since there are no external
    gluons at the Born level.};
 
\item  the Born phase space;
  
\item \label{item:real} the squared real matrix elements for all relevant
  partonic processes;

\item \label{item:virtual} the finite part of the virtual corrections
  computed in dimensional regularization;

\item the Born colour structures in the limit of a large  number of colours.
\end{enumerateroman}
The symmetric colour correlated ${\cal B}_{ij}$ amplitudes, according to the
particle labelling used in the \POWHEGBOX{} ($q(1)\,
\bar{q}^\prime(2)\rightarrow W(3)\,b(4)\,\bar{b}(5)$), are given by
\begin{eqnarray}
&& {\cal B}_{ii}=-\CF {\cal B}, \qquad i=1,2,4,5
\\
&&{\cal B}_{12}= \(\CF-\frac{\CA}{2}\){\cal B},\quad 
{\cal B}_{14}=\(2\CF-\frac{\CA}{2}\){\cal B},\quad
{\cal B}_{15}=  -\(2\CF-\CA\){\cal B},
\end{eqnarray}
for the case where $q$ is a quark. For the case where the first incoming
particle is an antiquark, we have the same results with ${\cal B}_{14}$ and
${\cal B}_{15}$ exchanged. Here $\CA=3$ and $\CF=4/3$ are the Casimir
invariants of the colour SU(3) representation.
We have taken the squared Born amplitude ${\cal B}$ from
ref.~\cite{FebresCordero:2006sj} while the squared real amplitudes have been
generated using MadGraph~\cite{Maltoni:2002qb}.

The assignment of colour flow for the two Feynman diagrams at the Born level
is straightforward and unambiguous, and follows directly the propagation of
quarks.

\subsection{Validation of the NLO code}

In the \POWHEGBOX{} framework, it is possible to compute NLO
distributions, taking advantage of the fact that the \POWHEGBOX{} computes
automatically all the counterterms needed to regularize the real
distributions.  It is then possible to check that, in the collinear and soft
limits, the real amplitude has the correct behaviour, and that it is
consistent with the Born cross section and the colour- and spin-correlated Born
amplitudes.

The NLO distributions obtained within the \POWHEGBOX{}, using a variant of
the FKS subtraction scheme~\cite{Frixione:1995ms}, have been checked against
the codes developed in refs.~\cite{FebresCordero:2006sj, Cordero:2008ce,
  Cordero:2009kv} which use instead a phase-space slicing method with two
cutoffs, to extract soft and collinear singularities
analytically~\cite{Harris:2001sx, Reina:2001bc, Dawson:2003zu}, as well as
independent calculations of both virtual and real corrections, performed with
several tools based on {\tt FORM}~\cite{Vermaseren:2000nd}, {\tt
  TRACER}~\cite{Jamin:1991dp} and {\tt MAPLE} codes.  Full agreement has been
found in all the studied distributions, assuring us that the ingredients
provided to the \POWHEGBOX{} are correct and consistent.

\section{Theoretical introduction}
\label{sec:theoretical_intro}
In the \POWHEG{} formalism, the generation of the hardest emission is
performed first, at full NLO accuracy, and subsequent radiation is generated
using shower Monte Carlo programs (see ref.~\cite{Frixione:2007vw} for a more
detailed description of the method).  In the following, we will briefly
summarize a few features of the \POWHEG{} method that will be useful in view
of the discussion of sec.~\ref{sec:mass_effects}.

\subsection{Generation of the underlying Born configuration}
The first step of the generation process is the construction of the
underlying Born kinematics, i.e.~the generation of the Born momenta,
distributed according to the function
\begin{equation}
\bar{B}(\Kinn) = B(\Kinn)+V(\Kinn)+\int d\Rad \, R(\Kinnpo)\,,
\end{equation}
where $B$, $V$ and $R$ are the Born, virtual and real contributions to the
NLO cross section, $\Kinn$ specifies the kinematics of the underlying Born
event with $n$ final-state particles, $\Rad$ are the radiation variables and
the real-emission variables $\Kinnpo\equiv\lg \Kinn,\Rad\rg$ are parametrized
in terms of the underlying-Born and radiation ones.

\subsection{Generation of the radiation variables}
Once the momenta of the underlying Born have been generated, the \POWHEG{}
method proceeds to generate the radiation, i.e.~$\Rad$, starting from the
\POWHEG{} cross section for the generation of the hardest emission
\begin{equation}
\label{eq:POWHEGsigmasimple}
d\sigma= \bar{B}(\Kinn)\, d \Kinn \Bigg\{ \Delta\(\Kinn,\ptmin\) +
\Delta\(\Kinn,\kt\(\Kinnpo\)\)\, \frac{ R\(\Kinnpo\)}{ B\!\(\Kinn\)} d\Rad
\Bigg\}\,,
\end{equation}
where values of $\kt\(\Kinnpo\)<\ptmin$ are not allowed
(here $\ptmin\sim 1$~GeV, i.e.~of the order of a typical hadronic
scale).  The Sudakov form factor $\Delta$ is given by
\begin{equation}
\label{eq:sudakov}
\Delta\(\Kinn, \pt\)=\exp\lg - \int d\Rad\,  \frac{R(\Kinnpo) \;
    }{B(\Kinn)}\,\stepf\!\(\kt\(\Kinnpo\)-\pt\) \rg\,.
\end{equation}
The function $\kt\(\Kinnpo\)$ should be equal, near the singular
limit, to the transverse momentum of the emitted parton relative to
the emitting one.  The cross section in
eq.~(\ref{eq:POWHEGsigmasimple}) has the following properties:
\begin{itemize}
\item at large $\kt$ it coincides with the NLO cross section up to
  next-to-next-to-leading order terms.  In fact, in the large
  transverse-momentum region, eq.~(\ref{eq:POWHEGsigmasimple}) can be written
  as
\begin{equation}
\label{eq:POWHEGsigmasimple1}
d\sigma= \bar{B}( {\bf \Phi}_n)\, \frac{ R\({\bf \Phi}_{n+1}\)}{
  B\!\({\bf \Phi}_{n}\)} \,d \Kinn \, d\Rad\,,
\end{equation}
since the Sudakov form factor approaches 1 in this region.  This
differs from the pure NLO result because of the presence of the factor
\begin{equation}
\label{eq:Bbar_over_B}
 \frac{ \bar{B}(\bf \Phi_n) }{ B\!\(\bf \Phi_{n}\)}=1+{\cal O}(\as)\;.
\end{equation}
For processes that get large radiative corrections, the ${\cal O}(\as)$ term
can be in fact of order 1, giving then a harder spectrum for the generated
radiation.

\item It reproduces correctly the value of infrared-safe observables at NLO.
  Thus, also its integral around the small $\kt$ region has NLO accuracy.
\item At small $\kt$ it behaves no worse than standard shower
  Monte Carlo generators.
\end{itemize}

\subsection{Tuning of the real contribution}
\label{sec:tuning_R}
In \POWHEG{} it is possible to tune the contribution to the real cross
section that is treated with the Monte Carlo shower technique.  This was
pointed out first in ref.~\cite{Nason:2004rx}, where the \POWHEG{} method was
formulated, and then implemented in the \POWHEGBOX{} as a general feature. In
this way, the enhancement in eq.~(\ref{eq:Bbar_over_B}) can be controlled, if
necessary. In fact, the real cross section can be split into two positive
contributions
\begin{equation}
\label{eq:singplusreg1}
  R = R_s + R_f\,, 
\end{equation}
such that $R_f$ has no singularities (soft or collinear) and only
$R_s$ is singular in the infrared regions. In previous
implementations~\cite{Alioli:2008tz} the separation was achieved using
a function $F$ of the transverse momentum of the radiation, $0
\leqslant F \leqslant 1$, that approaches 1 when the transverse
momentum of the radiated parton vanishes, and such that
\begin{eqnarray}
\label{eq:Rs}
  R_s & = & R \, F\,, \\
\label{eq:Rf}
  R_f & = & R \lq 1 - F \rq \,. 
\end{eqnarray}
The generation of the radiation is then done by \POWHEG{} using only the
divergent contribution $R_s$ and eqs.~(\ref{eq:POWHEGsigmasimple})
and~(\ref{eq:sudakov}) become
\begin{eqnarray}
\label{eq:POWHEGsigmasimple_s}
&&d\sigma = \bar{B}_s(\Kinn)\, d \Kinn \Bigg\{ \Delta_s\(\Kinn,\ptmin\) +
\Delta_s\(\Kinn,\kt\(\Kinnpo\)\)\, \frac{ R_s\(\Kinnpo\)}{ B\!\(\Kinn\)} d\Rad
\Bigg\} \nonumber \\
&& \phantom{d\sigma = } + R_f\(\Kinnpo\) d \Kinn\, d\Rad\,,
\\
&&
\label{eq:Bbar}
\bar{B}_s(\Kinn) = B(\Kinn)+V(\Kinn)+\int d\Rad \, R_s(\Kinnpo)\,,\\
&&
\label{eq:sudakov_s}
\Delta_s\(\Kinn, \pt\) = \exp\lg - \int d\Rad\,  \frac{R_s(\Kinnpo) \;
    }{B(\Kinn)}\,\stepf\!\(\kt\(\Kinnpo\)-\pt\) \rg\,.
\end{eqnarray}
The contribution $R_f$, being finite and positive, is generated with
standard NLO techniques, and fed into a shower Monte Carlo as is.

\subsection[$Wb\bar{b}$ production]{$\boldsymbol{Wb\bar{b}}$ production}
\label{sec:mass_effects}

\begin{figure}[htb]
\begin{center}
\epsfig{file=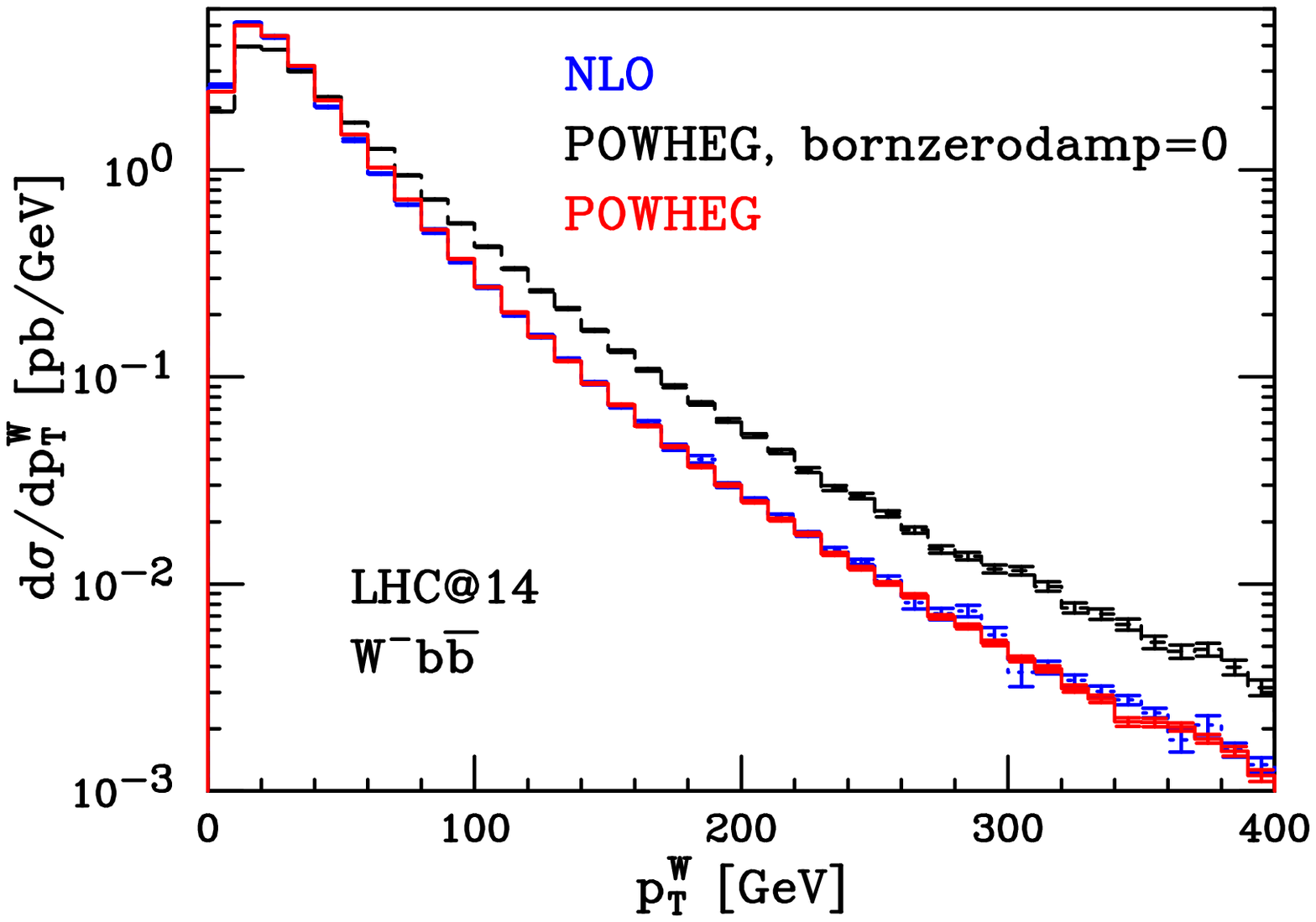,width=\wfig,height=\hfig}
\epsfig{file=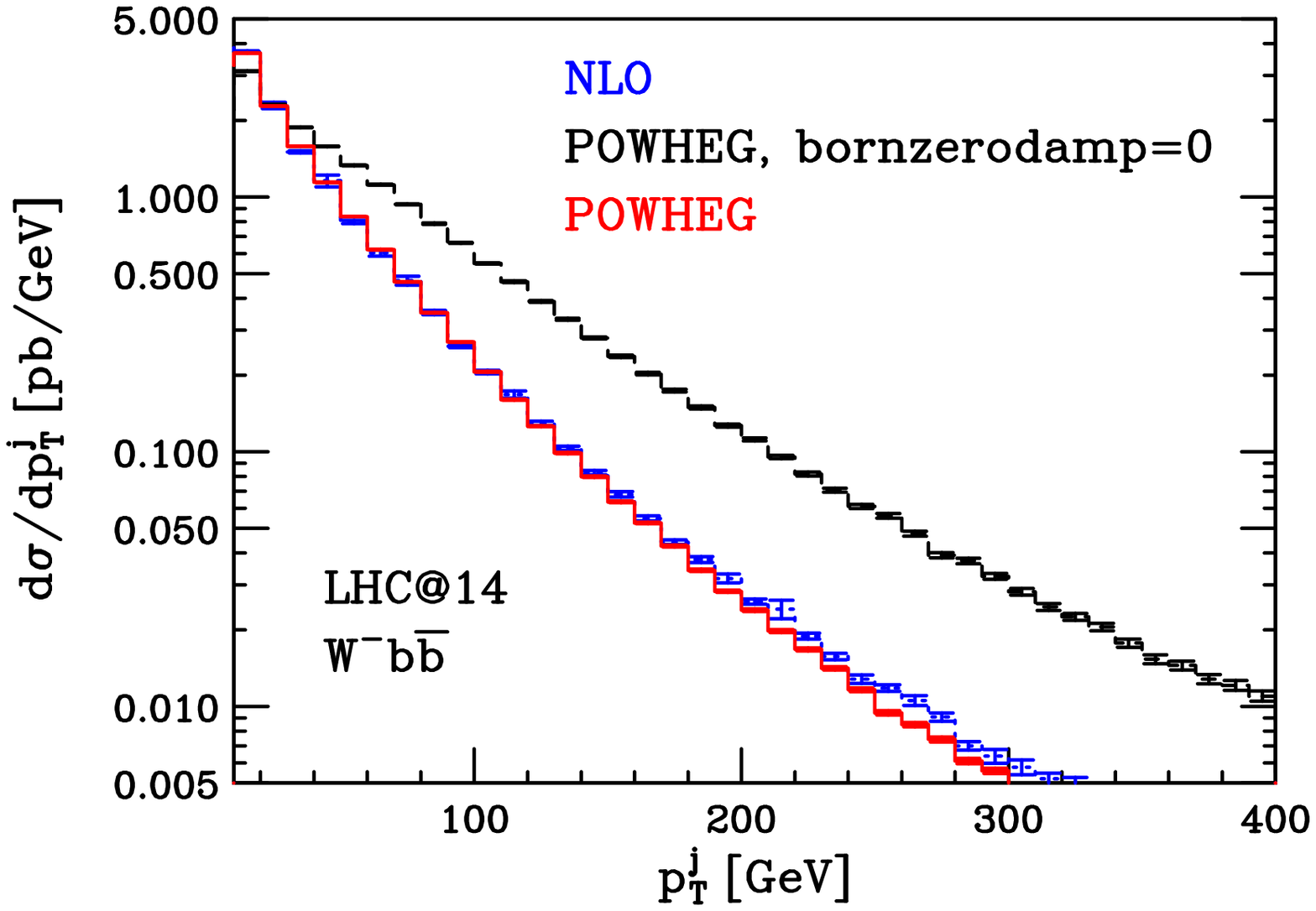,width=\wfig,height=\hfig}

\epsfig{file=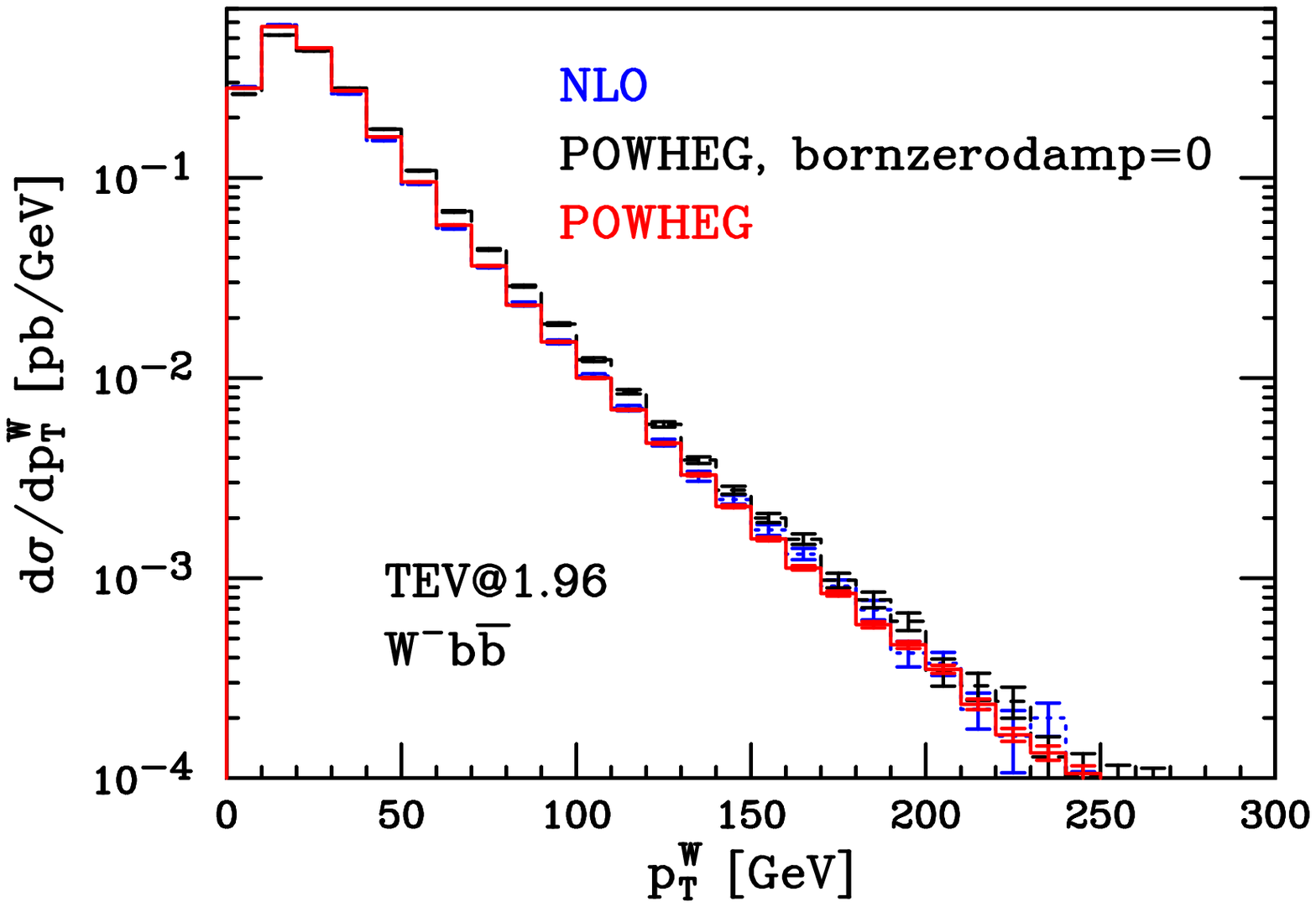,width=\wfig,height=\hfig}
\epsfig{file=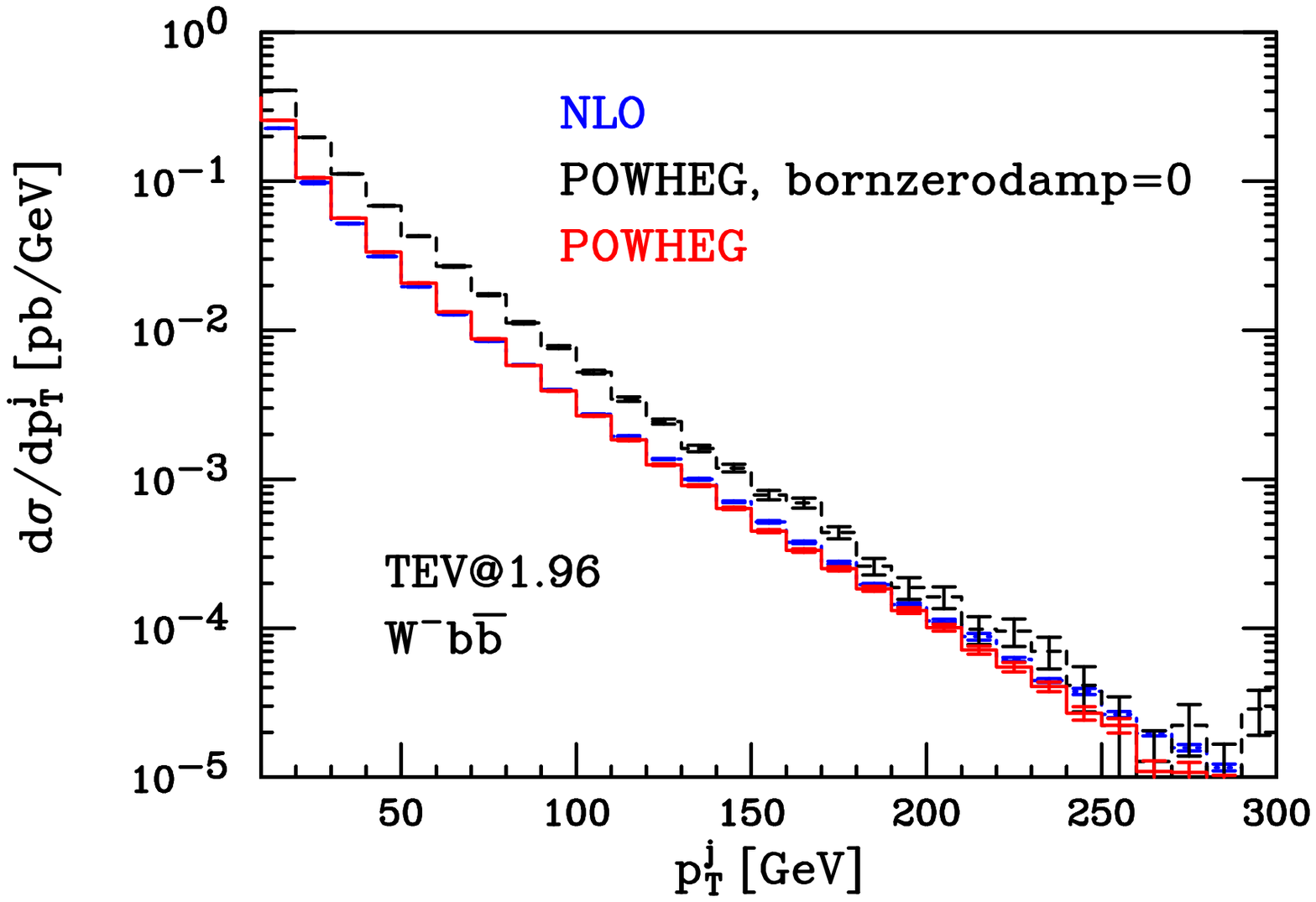,width=\wfig,height=\hfig}
\end{center}
\caption{\label{fig:LHC_TEV_ptj_ptw} Transverse-momentum distributions for
  the $W$ boson, $p_T^W$, and the hardest radiated non-$b$ jet, $p_T^j$, in
  $W^-b\bar{b}$ production at NLO in QCD, for both the LHC with
  $\sqrt{s}=14$~TeV (upper plots) and the Tevatron (lower plots). The dotted
  blue lines represent the results of the pure NLO QCD calculation.  The
  \POWHEG{} results obtained with the mechanism to protect from Born zero not
  activated and with no separation of the hard-radiation region are shown in
  dashed black lines.  In solid red, the results with the mechanism to
  protect from Born zero
  activated and with the separation of the hard radiation collinear to the bottom
  quarks, as described in sec.~\ref{sec:mass_effects}.}
\end{figure}

As for Higgs-boson production in gluon fusion~\cite{Alioli:2008tz}, also in
$Wb\bar{b}$ production the NLO corrections are very large, independently of
the choice of the renormalization and factorization
scale~\cite{FebresCordero:2006sj,Cordero:2009kv}. For example, using a
renormalization and factorization scale equal to $\mu=m_W+2m_b$ and input
parameters as specified in sec.~\ref{sec:phenomenology}, for $W^-b\bar{b}$
production at the LHC with $\sqrt{s}=14$~TeV, the LO cross section
corresponding to the distributions shown in fig.~\ref{fig:LHC_TEV_ptj_ptw} is
81.32~pb, while the NLO cross section is 222.88~pb, with a $K$ factor (ratio
of NLO over LO total cross section) of 2.74, while at the Tevatron, the cross
sections are 10.43~pb at LO, and 19.84~pb at NLO, with a $K$ factor of 1.9.

\begin{table}[htb]
\begin{center}
\begin{tabular}{|c|c|c|c|} 
\hline
$m_b$ [GeV] & LO [pb] & NLO [pb] & $K$ \\
\hline
0.1 & $140.6\pm 0.3$ & $412\pm 8$ & 2.9 \\
1.0 & $135.5\pm 0.2$ & $381\pm 2$ & 2.8 \\
10  & $37.11\pm 0.03$ & $98.2\pm 0.3$ & 2.6\\
100 & $0.5240\pm 0.0003$ & $0.961\pm 0.003$ & 1.8\\
\hline
\end{tabular}
\end{center}
\caption{\label{tab:Kfactors} Values of the LO and NLO cross sections, as
  well as their respective $K$ factors, for $W^-b\bar{b}$ production at the
  LHC with 14~TeV, using $\mu=m_W+2m_b$ as renormalization and factorization
  scale and varying the quark mass $m_b$ by three orders of magnitude.}
\end{table}

 
We can attribute these large NLO corrections to at least two reasons: the
opening of a new gluon-initiated channel in the real contributions at NLO
(e.g.~$qg\rightarrow Wb\bar{b}+q$), that is likely to be more important at
the LHC than at the Tevatron~\cite{FebresCordero:2006sj, Cordero:2008ce,
  Cordero:2009kv}, and the presence of large logarithms of the mass of the
bottom quark, related to final-state collinear singularities for massless
quarks, now regularized by the quark mass.  To illustrate this, we have
collected in table~\ref{tab:Kfactors} the $K$ factors for several values of
the mass of the final-state heavy quark, for the LHC at 14~TeV, and using
renormalization and factorization scales equal to $\mu=m_W+2m_b$. One can see
that the $K$ factors are large and, when the quark mass increases from
0.1~GeV to 100~GeV, the $K$ factor decreases from 2.9 to 1.8.
We expect the large collinear logarithms to affect distributions even more
than total cross sections.  

Before presenting a few results for $Wb\bar{b}$ production, we would like to
address a further issue related to this process: $Wb\bar{b}$ production has a
Born zero when a relativistic $W$ is emitted parallel with respect to the
incoming beam, and consequently the gluon (which undergoes the splitting into
a $b\bar{b}$ pair) is emitted in the opposite direction.  Along the incoming
beam, angular-momentum conservation cannot be preserved, and this
configuration is kinematically suppressed. This was in fact the case for $W$
production, where there is a zero in the Born cross section if the outgoing
lepton from $W$ decay is anti-parallel to the incoming
quark~\cite{Alioli:2008gx}. In fact, due to the left-handed nature of the $W$
boson coupling, we have a violation of angular-momentum conservation along
the incoming beam.

Despite the fact that $B({\bf \Phi}_{n}^0)$ can vanish in some kinematic
regions ${\bf \Phi}_{n}^0$, this kinematics can be generated since the
$\bar{B}_s({\bf \Phi}_{n}^0)$ term in eq.~(\ref{eq:Bbar}) can be different
from zero, due to the real term.  From eq.~(\ref{eq:POWHEGsigmasimple_s}), we
see that, in this case, away from the Sudakov region (i.e.~where
$\Delta_s\approx 1$), the real contribution may be enhanced by a factor
$\bar{B}_s/B$, now large since $B\to 0$.
The \POWHEGBOX{} has a built-in mechanism to deal with this problem: $R^s$ is
chosen to vanish in the regions where $R$ differs too much (by more than a
factor of 5) from its collinear or soft approximation, which are proportional
to the underlying Born cross section. The contribution $R_f=R-R_s$, being
non-singular, is then added independently.  To activate this mechanism, the
\POWHEGBOX{} flag {\tt bornzerodamp} has to be set to 1. This must then
necessarily be done for the process we are studying.

In fig.~\ref{fig:LHC_TEV_ptj_ptw}, we have plotted the transverse momentum of
the $W$ boson and of the hardest radiated jet at the Tevatron and at the
LHC. Jets are reconstructed using the anti-$\kt$
algorithm~\cite{Cacciari:2008gp} with $R=0.4$, and jets are recombined using
the default recombination $E$ scheme~\cite{Catani:1993hr}. No other cuts are
applied to the events. The values of all the coupling constants, masses and
physical parameters for the generation of these events can be found in
sec.~\ref{sec:phenomenology}.  In dashed black lines, the \POWHEG{}
hardest-emission results when the {\tt bornzerodamp} flag is set to {\tt false}.
It is evident the effect of the enhancing factor $\bar{B}_s/B$ with respect
to the corresponding NLO results (dotted blue lines).

In addition to the activation of the mechanism to protect from the Born zeros,
we have decided to separate out the region of hard-gluon emission collinear
to one of the final-state massive quarks from the part of the real
contribution that is treated by the Monte Carlo shower techniques
(i.e.~generated through the Sudakov form factor), and to handle it with
standard NLO techniques, as described in sec.~\ref{sec:tuning_R}.  In fact,
this region is responsible for the presence of large logarithms of the mass
of the quark and we do not want these logarithmic terms to get further
enhanced by the $\bar{B}/B$ ratio of eq.~(\ref{eq:Bbar_over_B}).  We would
like to stress that, strictly speaking, this region is not singular,
since the mass of the quark regularizes it, but it would be a singular region
if the mass of the quark were exactly zero.

In order to separate out the region of hard-gluon emission collinear to the
final-state $b$ quarks, we have chosen the following form for the function
$F$ of eqs.~(\ref{eq:Rs}) and~(\ref{eq:Rf})
\begin{equation}
\label{eq:funct_F}
F= \frac{\(1/d\)^c}{\(1/d\)^c + \(1/d_b\)^c + \(1/d_{\bar{b}}\)^c}\,,
\end{equation}
where 
\begin{eqnarray}
d &=& E^2\(1-\cos^2\theta\)\,, \\
d_b &=& \frac{ E \, E_b}{(E+E_b)^2}  \frac{(E+m_b)^2}{E^2} \, k\cdot k_b  =
E_b^2\,  \frac{(E+m_b)^2 }{(E+E_b)^2}  \(1 -
 \frac{|\vec{k}_b|}{E_b}\cos\theta_b\)  \,,
\end{eqnarray}
and where $k$ is the momentum of the radiated gluon with energy $E$, forming
an angle $\theta$ with the positive direction of the incoming beam, and an
angle $\theta_b$ with the outgoing $b$ quark, in the center-of-mass frame.
The momentum of the $b$ quark is $k_b$, with energy $E_b$ and three-momentum
$\vec{k}_b$. The $d_{\bar{b}}$ term is similar to the $d_b$ one except for
exchanging $b$ with $\bar{b}$ in all the kinematic variables.  We have set
$c=1$ in the code, but higher values can be used too.

The function $F$ in eq.~(\ref{eq:funct_F}) has the following properties:
\begin{enumerate}
\item it approaches 1 in the singular region, i.e.~when the emitted gluon is
  parallel to the incoming beams or soft ($d\to 0$), assuring that the
  singular region is treated with the Monte Carlo shower technique;

\item it becomes small when the radiated parton is hard and collinear to the
  $b$ or the $\bar{b}$ quark. In fact, when the radiated gluon is hard and
  collinear to one of the two heavy quarks, the $d_b$ or $d_{\bar{b}}$
  terms reach their minimum value.

\end{enumerate}
The distributions obtained with the mechanism to protect from the Born zeros
and with the $F$ function of eq.~(\ref{eq:funct_F}) are plotted as solid red
lines in fig.~\ref{fig:LHC_TEV_ptj_ptw}. We find very good agreement with the
NLO curves, at least in the region where we expect this to happen. The
expected disagreement in the low-$\pt$ jet region is due to the fact that the
NLO curve is divergent in this region, while the resummed \POWHEG{} result
feels the effect of the Sudakov form factor and goes correctly to zero.  In
addition, we have verified that the major role in getting this agreement is
played by the activation of the mechanism to protect from Born-zero
configurations, while the effect of the separation of the
region of hard gluons parallel to the massive $b$ quarks plays only a minor
role.

\subsection[$W$-boson decay]{${\boldsymbol W}$-boson decay}
\label{sec:w_decay}
Since we have used the analytic calculation of the virtual corrections of
refs.~\cite{FebresCordero:2006sj, Cordero:2008ce, Cordero:2009kv}, that
treats the $W$ boson as stable, we are not in a position to have all the spin
correlations in the leptonic $W$-boson decay products correctly accounted
for. We can instead use standard techniques to implement the decay in an
approximated way~\cite{Frixione:2007zp}.  In this approximation, spin
correlations are not accurate to NLO in the whole phase space, but are
correct to NLO for hard real emissions and to leading order in the soft and
collinear region.
In order to achieve this, we have produced a $W$ boson with invariant mass
$M$ distributed according to the Breit-Wigner function
\begin{equation} 
\frac{1}{\pi} \frac{m_W\Gamma_W}{(M^2-m_W^2)+m_W^2\Gamma_W^2}\,,
\end{equation} 
where $m_W$ and $\Gamma_W$ are the pole mass and width of the $W$ boson.
Using the \POWHEG{} method, a Born-like, or real-like event, is generated
with an undecayed $W$ boson, whose invariant mass is $M$, and whose
kinematics is parametrized by a set of variables that we call $\Phi_{\rm u}$,
where ``u'' stands for ``undecayed''.

The procedure that we are going to follow can be easily illustrated if we
recall that the squared matrix elements are connected to the concept of
probability.  We then rephrase the procedure that we have used in a
probabilistic language.  The differential probability distribution of the
decay variables is proportional to (we neglect the overall normalization
factor that ensures that the integral of the differential probability
distribution is 1)
\begin{equation}
\label{eq:sigma_dec}
dP(\Phi_{\rm d}) \div {\cal M}_{\rm d}(\Phi_{\rm d})\,d\Phi_{\rm d} =
{\cal M}_{\rm d}(\Phi_{\rm u},\Phi_{W\rightarrow l\nu}) \,
d\Phi_{\rm u}\, d \Phi_{W\rightarrow l\nu}\,,
\end{equation}
where ${\cal M}_{\rm d}$ is the squared amplitude corresponding to the
decayed process, with finite-width effects fully taken into
account\footnote{The decayed tree-level squared amplitudes ${\cal M}_{\rm
    d}$ have been obtained using MadGraph~\cite{Maltoni:2002qb}, both for the
  Born and for the radiative processes.}.  For consistency, the squared
amplitude ${\cal M}_{\rm d}$ must include only resonant diagrams
(i.e.~diagrams where the $W$ momentum equals the sum of the $ l$ and $\nu$
momenta).  In writing eq.~(\ref{eq:sigma_dec}), we have parametrized the
kinematics of the process for the production and decay of the $W$ boson in
terms of the undecayed variables $\Phi_{\rm u}$ and of a set of variables
describing the $W$ decay, $\Phi_{W\rightarrow l\nu}$.
Equation~(\ref{eq:sigma_dec}) implicitly defines $d \Phi_{W\rightarrow
  l\nu}$.  Similarly, the differential probability distribution of the
undecayed variables is proportional to ${\cal M}_{\rm u}(\Phi_{\rm u})\,
d\Phi_{\rm u} $. The problem is then to determine the probability
distribution of the variables that parametrize the decay, $\Phi_{W\rightarrow
  l\nu}$, given the probability distributions for $\Phi_{\rm d}$ and
$\Phi_{\rm u}$. To solve this problem we use the fact that the joint
probability of two events $A$ and $B$ can be written in terms of the
conditional probability
\begin{equation} 
P(A \cup B) = P(A|B) \times P(B)\,,
\end{equation}
that in our case becomes ($A$ corresponds to the generation of the decay
variables for $W\to l\nu$ and $B$ corresponds to the generation of the
undecayed $\Wbb$ event)
\begin{equation}
\label{eq:cond_probab}
{\cal  M}_{\rm d}(\Phi_{\rm d})\, d\Phi_{\rm d} \div
dP(\Phi_{W\rightarrow l\nu} \,|\, \Phi_{\rm u}) \times
{\cal  M}_{\rm u}(\Phi_{\rm u})\, d\Phi_{\rm u}\,,
\end{equation}
where $dP(\Phi_{W\rightarrow l\nu}\,|\, \Phi_{\rm u}) $ is the infinitesimal
probability distribution of the variables $\Phi_{W\rightarrow l\nu}$,
given the kinematics of an undecayed process $\Phi_{\rm u}$.
We can then write, using eq.~(\ref{eq:sigma_dec}),
\begin{eqnarray}
dP(\Phi_{W\rightarrow l\nu} \,| \,\Phi_{\rm u})  &\div& 
\frac{{\cal  M}_{\rm d}(\Phi_{\rm d})\, d\Phi_{\rm d}}{{\cal  M}_{\rm
    u}(\Phi_{\rm u})\, d\Phi_{\rm u}} =
\frac{{\cal M}_{\rm d}(\Phi_{\rm u},\Phi_{W\rightarrow l\nu}) \,
d\Phi_{\rm u}\, d \Phi_{W\rightarrow l\nu}}{{\cal  M}_{\rm
    u}(\Phi_{\rm u})\, d\Phi_{\rm u}}
\nonumber\\
&=&
\label{eq:diff_prob_decay}
\frac{{\cal M}_{\rm d}(\Phi_{\rm u},\Phi_{W\rightarrow l\nu})}
{{\cal  M}_{\rm    u}(\Phi_{\rm u})} \,
 d \Phi_{W\rightarrow l\nu}\,.
\end{eqnarray}
$dP(\Phi_{W\rightarrow l\nu} \,| \,\Phi_{\rm u})/d \Phi_{W\rightarrow l\nu}$
is the distribution function we are looking for.  To generate efficiently
$\Phi_{W\rightarrow l\nu}$, distributed according to
(\ref{eq:diff_prob_decay}), we use the hit-and-miss technique and so we need
to find an upper bound for the ratio ${{\cal M}_{\rm d}(\Phi_{\rm
    u},\Phi_{W\rightarrow l\nu})}/ {{\cal M}_{\rm u}(\Phi_{\rm u})}$. We have
used as upper bounding function the expression
\begin{equation}
U_{\rm d}(M^2,\Phi_{W\rightarrow l\nu})=N_{\rm d}\,
\frac{2M^2+m_l^2}
{(M^2-m_W^2)^2+m_W^2\Gamma_W^2}\,{\cal M}_{W\rightarrow l\nu}(M^2)\,,
\end{equation}
where $N_{\rm d}$ is a normalization factor, ${\cal M}_{W\rightarrow l\nu}$
is the squared decay amplitude corresponding to the $W\rightarrow l\nu$ decay
and $m_l$ is the charged-lepton mass. One can predict the appropriate
value for the normalization factor $N_{\rm d}$ or compute it by sampling the
decay phase space $\Phi_{W\rightarrow l\nu}$ and comparing $U_{\rm d}$ with
the exact expression, in such a way that the inequality
\begin{equation}
\frac{{\cal M}_{\rm d}(\Phi_{\rm u},\Phi_{W\rightarrow l\nu})}
{{\cal  M}_{\rm    u}(\Phi_{\rm u})}  \le 
U_{\rm d}(M^2,\Phi_{W\rightarrow l\nu})
\end{equation}
holds. The veto algorithm is then applied as follows:
\begin{enumerate}
\item first one generates a point in the phase space $\Phi_{W\rightarrow
  l\nu}$;
\item then a random number $r$ in the range $[0,U_{\rm
    d}(M^2,\Phi_{W\rightarrow l\nu})]$ is generated;
\item finally, if $r< {{\cal M}_{\rm d}(\Phi_{\rm u},\Phi_{W\rightarrow
    l\nu})}/ {{\cal M}_{\rm u}(\Phi_{\rm u})}$, the kinematics of the decay
  is kept and the event is generated. Otherwise the algorithm goes back to
  step 1.
\end{enumerate}
At the end of this procedure, the kinematics $\Phi_{W\rightarrow l\nu}$ of
the $W$ decay is generated, and, together with the \POWHEG-generated
undecayed variables $\Phi_{\rm u}$, the kinematics of $Wb\bar{b}$ event
followed by the decay of the $W$ boson becomes available.

\begin{figure}[htb]
\begin{center}
\epsfig{file=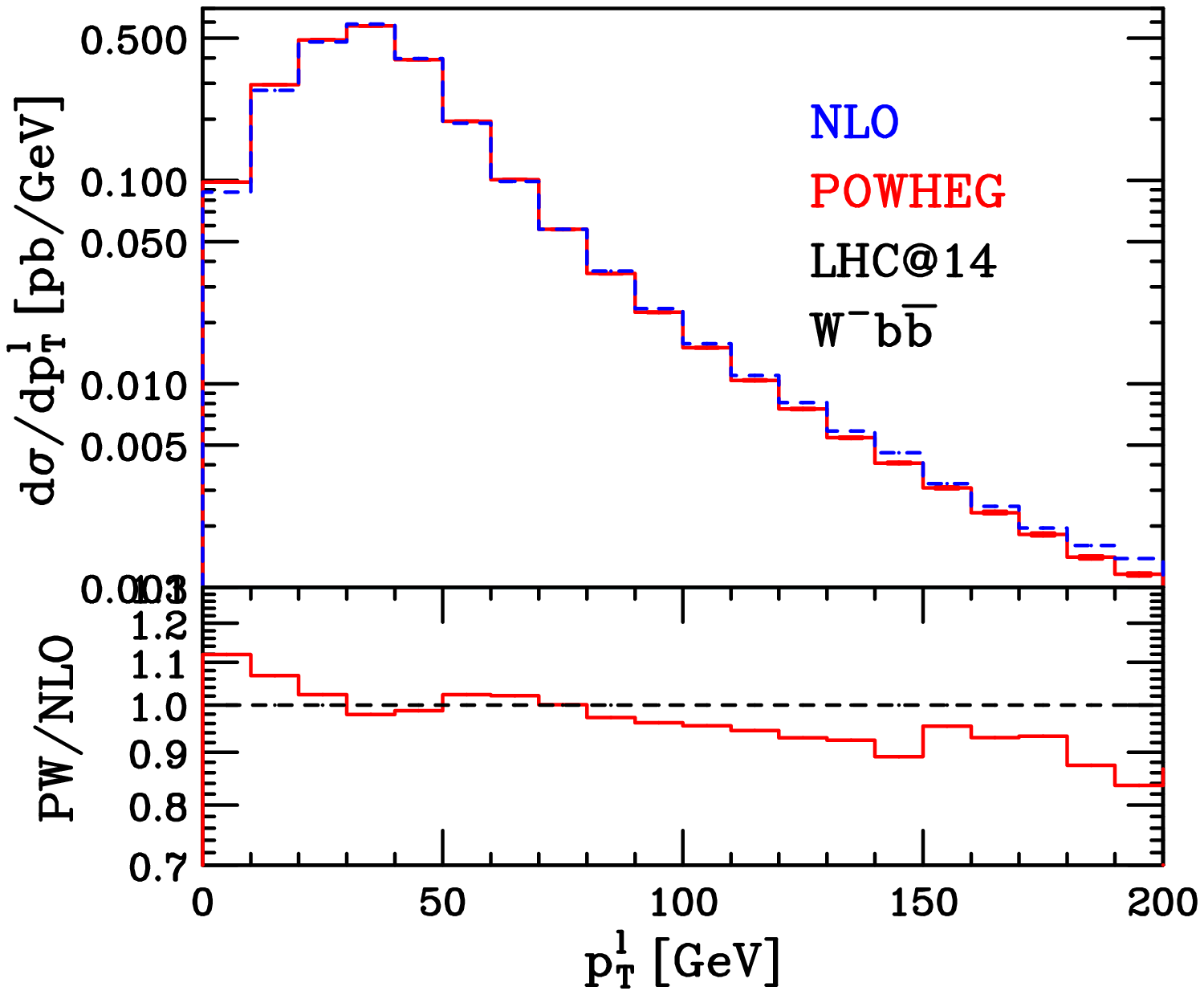,width=\wfig,height=\hfig}
\epsfig{file=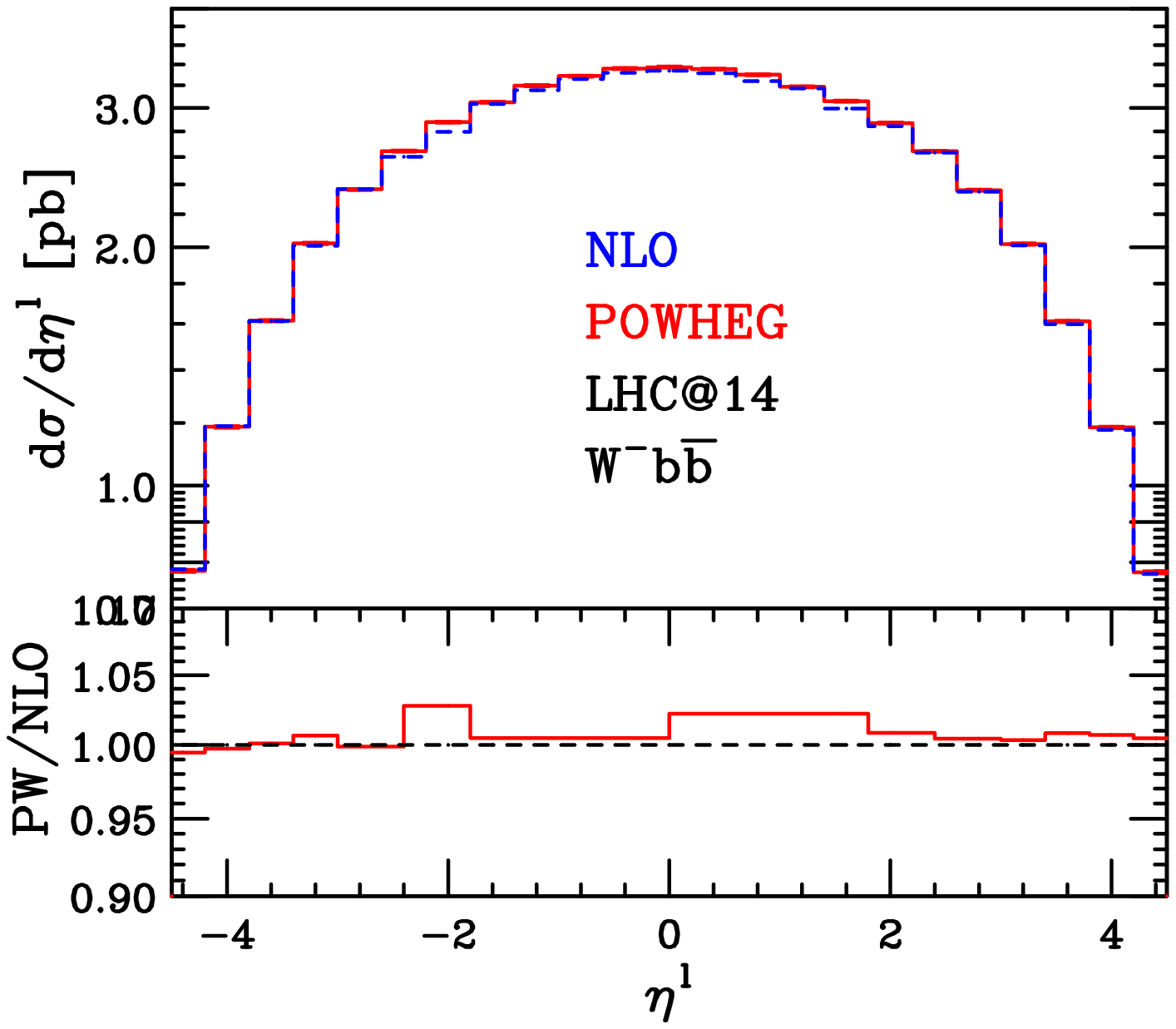,width=\wfig,height=\hfig}
\end{center}
\caption{\label{fig:LHC_lepton} Transverse momentum, $p_T^l$, and
  pseudorapidity, $\eta^l$, distributions for the final-state lepton produced
  in $\Wmbbdec$, at the LHC with $\sqrt{s}=14$~TeV. The distributions
  generated by the \POWHEGBOX{} according to the method described in
  sec.~\ref{sec:w_decay} are shown in solid red lines, while the NLO
  results computed with the \MCFM{} code are shown in dashed blue lines. The
  lower inserts show the ratio between the two distributions: {\tt
    POWHEG/NLO}.}
\end{figure}

In fig.~\ref{fig:LHC_lepton} we have plotted the differential cross section
as a function of the transverse momentum of the charged lepton $\pt^l$ and
its pseudorapidity $\eta^l$, generated with the procedure described above, in
solid red lines.  For comparison we have plotted, in dashed blue lines, the
NLO differential cross section for $\Wmbbdec$ production of
ref.~\cite{Badger:2010mg}, as implemented in \MCFM~6.0~\cite{MCFM}, where
spin correlations of the decay products have been included exactly.  The
branching ratio applied to the \POWHEG{} hardest-emission results has been
taken from the ratio
\begin{equation}
{\rm BR}(W \to l\nu) = \frac{\sigma_{\rm \sss LO, MCFM}}{\sigma_{\rm \sss LO, POWHEG}}=0.103\,.
\end{equation}
As can be seen from the insert in the lower part of the two plots, where we
have plotted the ratio of the \POWHEG{} hardest emission results over the
exact NLO ones ({\tt PW/NLO}), the approximated decay distributions are in
very good agreement with the exact ones.

\section{Phenomenology}
\label{sec:phenomenology}

In this section we present a few results for $\Wmbbdec$ production at the
Tevatron and at the LHC. Similar results can be obtained for $W^+ b\bar{b}\to
l^+ \nu\, b\bar{b}$.

For future reference, we list here the values of all the parameters and
physical quantities that enter the calculation:
\beqn
&&m_W=80.41~\GeV\,,\quad \quad m_{b} =4.62~\GeV\,, \quad \quad
m_t=173.1~\GeV\,, \nonumber\\
&&\Gamma_W=2.141~\GeV\,, \quad\quad  {\rm BR}(W \to l\nu) = 0.103\,,
\eeqn
and
\begin{equation}
\sin^2\theta_W = 0.223\,, \qquad \alpha = 1/132.088832\,, \qquad  G_F=1.16639\times
10^{-5}~{\GeV}^{-2}\,,
\end{equation}
from which we derive ($g_W^2=8m_W^2 G_F/\sqrt{2}$)
\begin{equation}
g_W=0.6532\,.
\end{equation}
We have used the CTEQ6.6 pdf set~\cite{Nadolsky:2008zw}, and we have set the
renormalization and factorization scale to the fixed value
\begin{equation}
\mu=m_W + 2\, m_b\,,
\end{equation}
from which we compute the two-loop $\overline{{\rm MS}}$ strong coupling constant
$\alpha_s(\mu)=0.1183$ with 5 light flavors.  The $W$-boson couplings to
quarks are proportional to the Cabibbo-Kobayashi-Maskawa~(CKM) matrix
elements.  We use non-zero CKM matrix elements for the first two quark
generations, $V_{ud}=V_{cs}=0.974$ and $V_{us}=V_{cd}=0.227$, while we
neglect the contribution of the third generation, since it is suppressed
either by the initial-state quark pdfs or by the corresponding CKM matrix
elements.

Jets are defined using the anti-$\kt$ algorithm~\cite{Cacciari:2008gp} with
$R=0.4$ and ${\kt}_{\min} = 5$~GeV, and are recombined using the default $E$
scheme.

Since there are no data for $W$ plus two $b$ jets to compare our predictions
with, and consequently no experimental analysis is available, we have chosen
a set of cuts that, while reasonable, are less stringent than the
experimental ones. In fact, the purpose of this section is to show the
differences between several \POWHEG{} results, obtained with different
showering programs, rather then provide predictions for experimentalists, who
can use the \POWHEGBOX{} by themselves to generate events and analyze them
according to their experimental selection criteria.

The set of cuts for the Tevatron, $\sqrt{s} = 1.96~\TeV$, and
the LHC, $ \sqrt{s} = 14~\TeV$, are the following:
\begin{eqnarray}
&&  \pt^b >15~\GeV\,, \qquad |\eta^b|< 3\,,\qquad  \pt^j > 15~\GeV\,,
\qquad |y^j| < 3\,, 
\nonumber \\
&& 
\label{eq:cuts}
\pt^l > 15~\GeV\,, \qquad |\eta^l|< 3\,,\qquad \etmiss > 15~\GeV\,.
\end{eqnarray}
We keep only events with at least two $b$-jets that pass the cuts on the
transverse momentum $\pt^b$ and on the pseudorapidity $\eta^b$, disregarding
all the other events.  Non-$b$ jets are required to have a minimum transverse
momentum of 15~\GeV{} and to be in the rapidity region $|y^j| < 3$.  Since we
have decayed the $W$ boson, we have cuts on the transverse momentum $\pt^l$
and pseudorapidity $\eta^l$ of the charged lepton, and a cut on the missing
energy $\etmiss$, due to the presence of the undetected neutrino.

In the following, we present several kinematic distributions both for the
Tevatron and the LHC, and we plot results for the hardest-emission
\POWHEG{} cross section with no shower (dotted black lines), and for the same
results showered by \PYTHIA{} (red solid curves) and \HERWIG{} (dashed blue
lines).  We have run \PYTHIA{} with the Perugia~0 tuning, switching off
multi-particle interactions\footnote{We have set {\tt mstp(81)=20} in the
  code, after the call to {\tt pytune}.}, in order to make a fair comparison
with \HERWIG{}, that uses the separate package {\tt
  JIMMY}~\cite{Butterworth:1996zw} to generate multi-particle
interactions. We have run \HERWIG{} in its default configuration, with
intrinsic $\pt$-spreading of 2.5~GeV.

Our analysis is based on a sample of 29 million events for the Tevatron and
26 million events for the LHC, generated with the \POWHEGBOX{} with no
folding (see ref.~\cite{Alioli:2010xd} for more details).  For the Tevatron
we got a 15\% fraction of negative-weight events and 12\% for the LHC, that
we have kept in our analysis.
\begin{table}[htb]
\begin{center}
\begin{tabular}{|c|c|c|c||c|c|c|c|}
\hline
$\xi$ &  $y$ & $\phi$ & negative fraction (\%)&$\xi$ &  $y$ & $\phi$ & negative fraction (\%)\\
\hline
1 & 1 & 1 & 12 &  2 & 5 & 10 & 2.3 \\
1 & 1 & 10 & 8 &  5 & 5 & 5 & 1.8\\
1 & 10 & 2 & 7 &  5 & 5 & 10 & 1.5\\
10 & 1 & 5 & 5 &  10 & 10 & 10 & 0.9\\
\hline
\end{tabular}
\end{center}
\caption{\label{tab:LHC_folding}
  Negative-weight fractions as a function of the folding in the radiation
  variables $\Rad=\{\xi, y,\phi\}$, at the LHC.}
\end{table}
If one is interested in the generation of only positive-weight events, then
the foldings of the radiation variables ($\Rad=\{\xi, y,\phi\}$) should be
increased.  We have collected in table~\ref{tab:LHC_folding} a few results
for different values of the foldings in the $\xi$, $y$ and $\phi$ variables.
From the table it is evident that, as the product of the folding numbers
increases, the fraction of negative-weight events gets smaller and smaller,
but at the price of a decrease in the speed of the code.

\subsection{Tevatron results}
\label{sec:phenomenolgy_tev}

\begin{figure}[htb]
\begin{center}
\epsfig{file=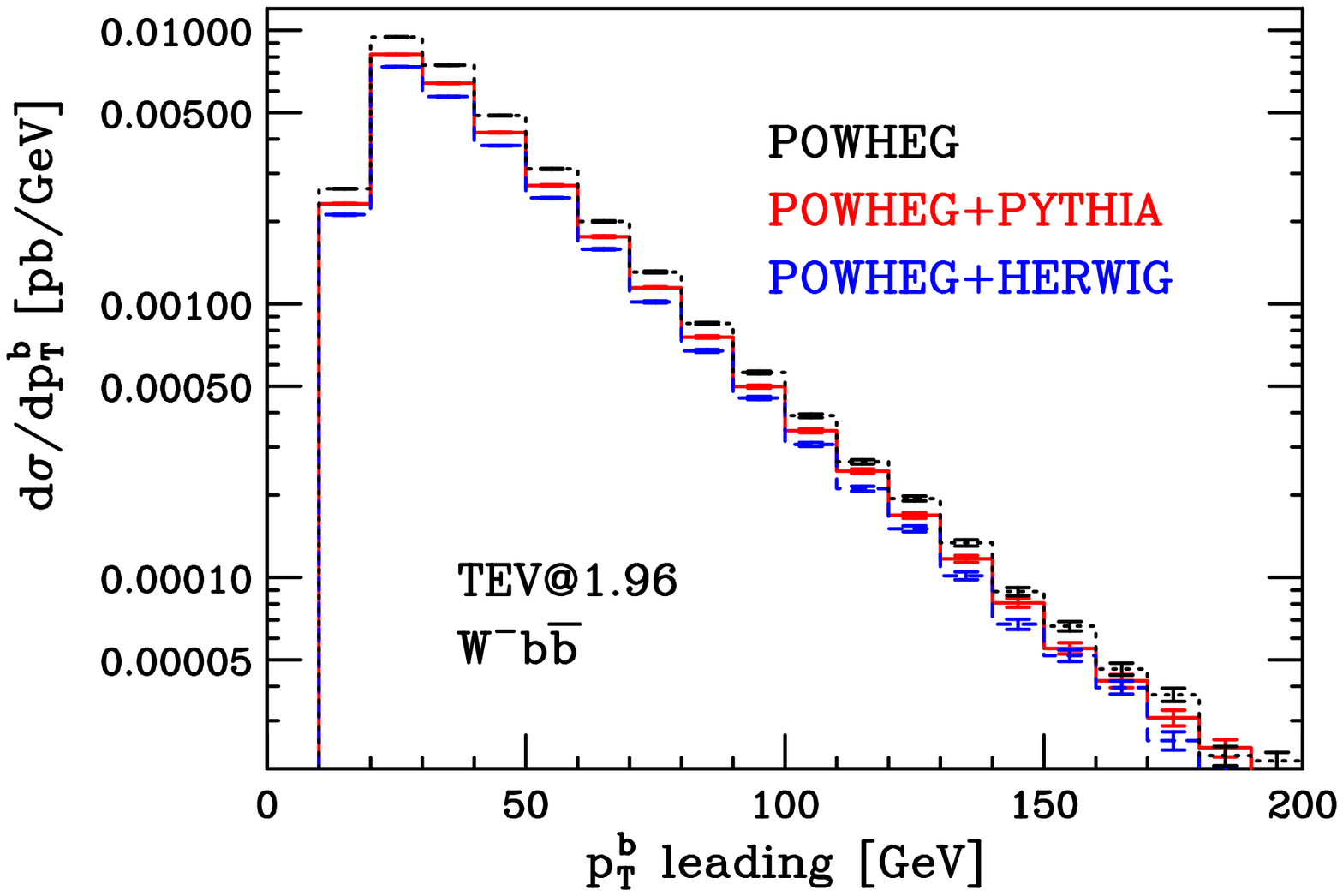,width=\wfig,height=\hfig}
\epsfig{file=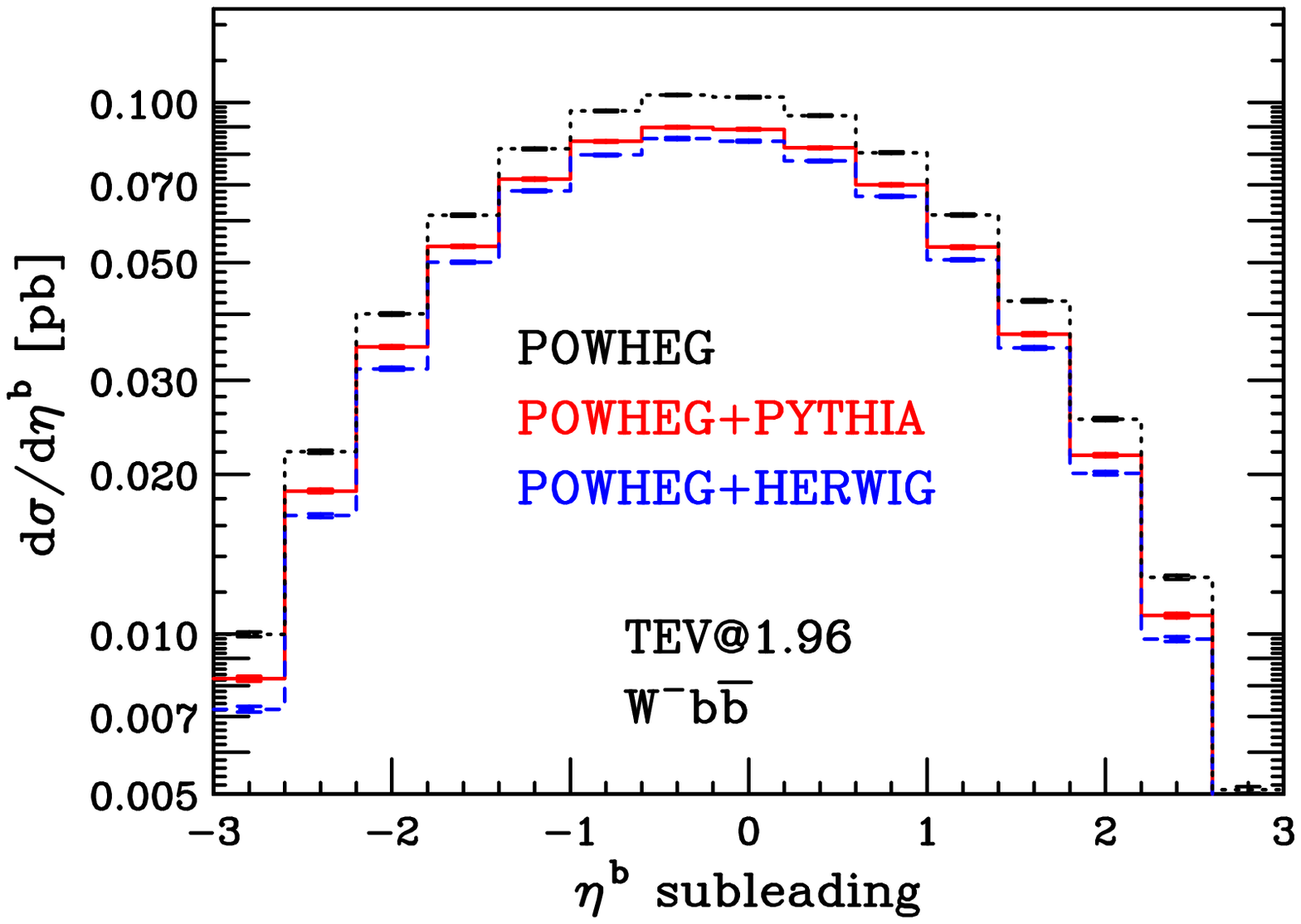,width=\wfig,height=\hfig}

\epsfig{file=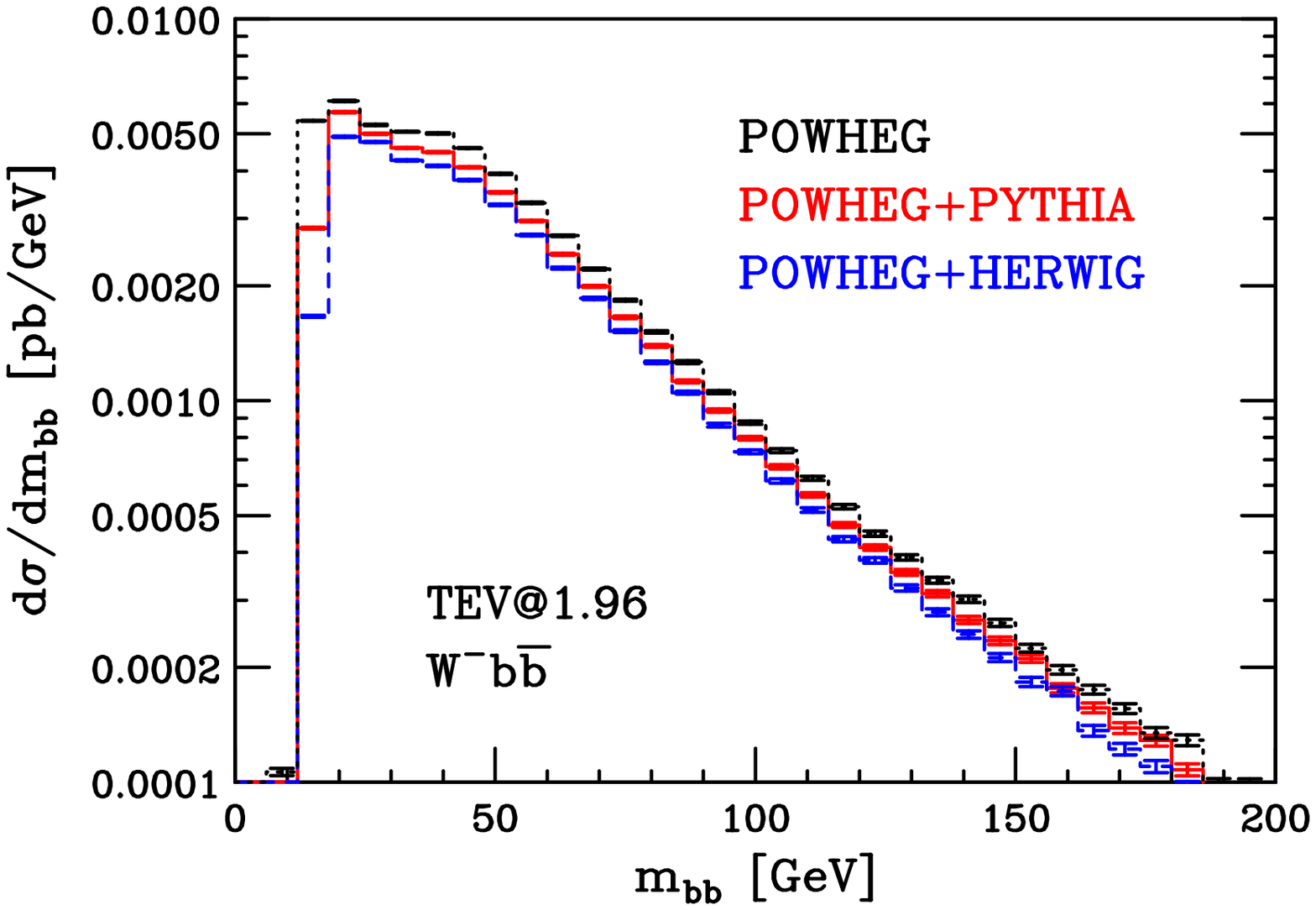,width=\wfig,height=\hfig}
\epsfig{file=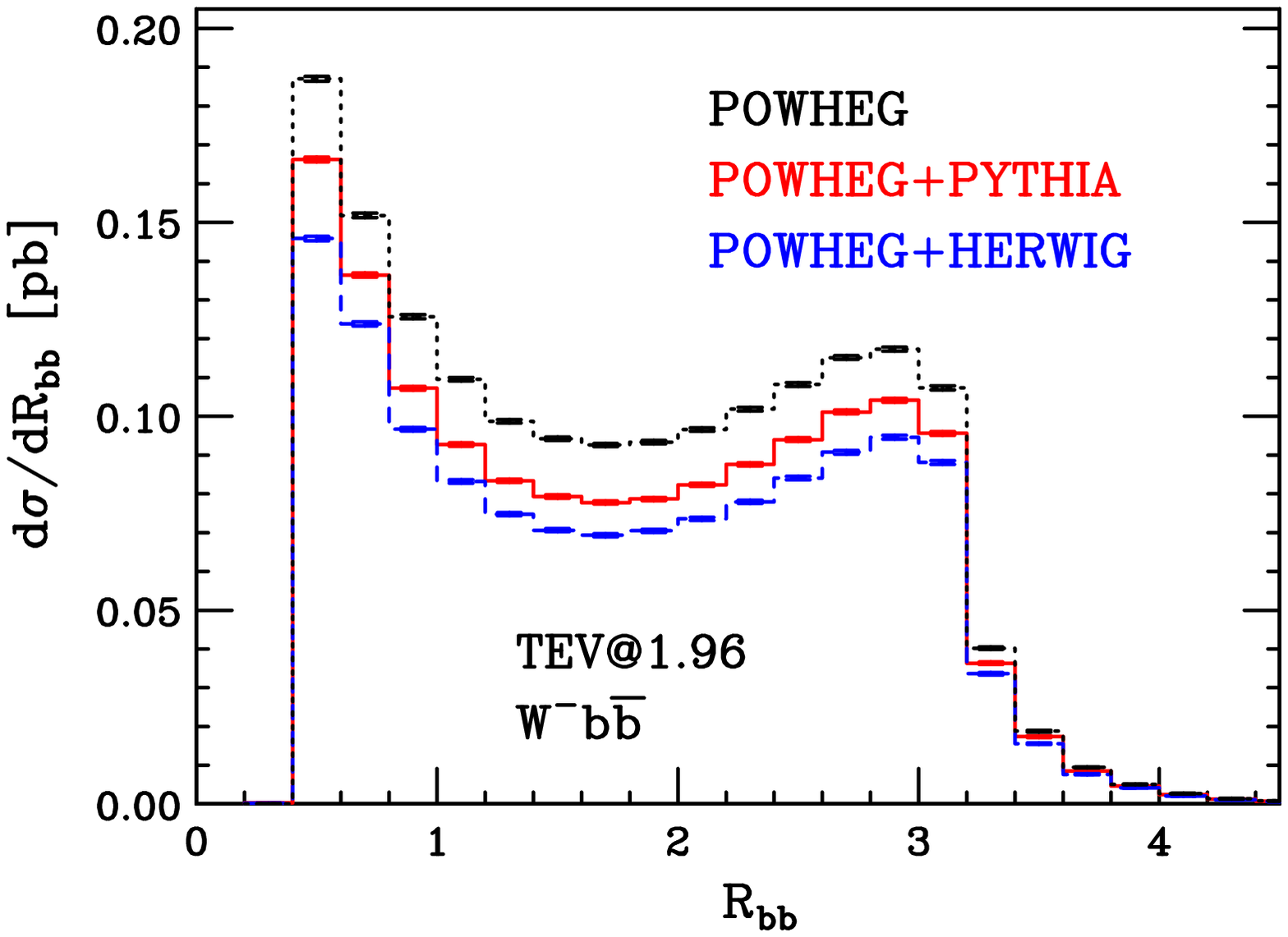,width=\wfig,height=\hfig}
\end{center}
\caption{\label{fig:TEV_PY_HW_1} Differential cross sections as a function of
  the transverse momentum of the hardest $b$ jet, $p_T^b$ leading, the
  pseudorapidity of the second hardest $b$ jet, $\eta^b$ subleading, the
  invariant mass of the leading and subleading $b$ jets, $m_{bb}$, and
  their angular distance $R_{bb}$, for $\Wmbbdec$ production at the
  Tevatron. The different curves represent the results for the \POWHEG{}
  hardest emission (dotted black), and for \POWHEG{} interfaced with either
  \PYTHIA{} (solid red) or \HERWIG{} (dashed blue).}
\end{figure}
In fig.~\ref{fig:TEV_PY_HW_1} we have plotted the differential cross sections
as a function of the transverse momentum of the $b$ jet with the hardest
$\pt$ (called leading $b$ jet), of the pseudorapidity of the $b$ jet with the
second hardest transverse momentum (called subleading $b$ jet), of the
invariant mass of the leading and subleading $b$ jets and of the angular
distribution $R_{bb}$, defined as
\begin{equation}
R_{bb} = \sqrt{\Delta y_{bb}^2 + \Delta\phi_{bb}^2}\,,
\end{equation}
where $\Delta y_{bb}$ and $\Delta\phi_{bb}$ are the difference in the
rapidities and in the azimuthal angles of the two highest-$\pt$ $b$ jets.
Error bars from the Monte Carlo generator are shown too.

\begin{figure}[htb]
\begin{center}
\epsfig{file=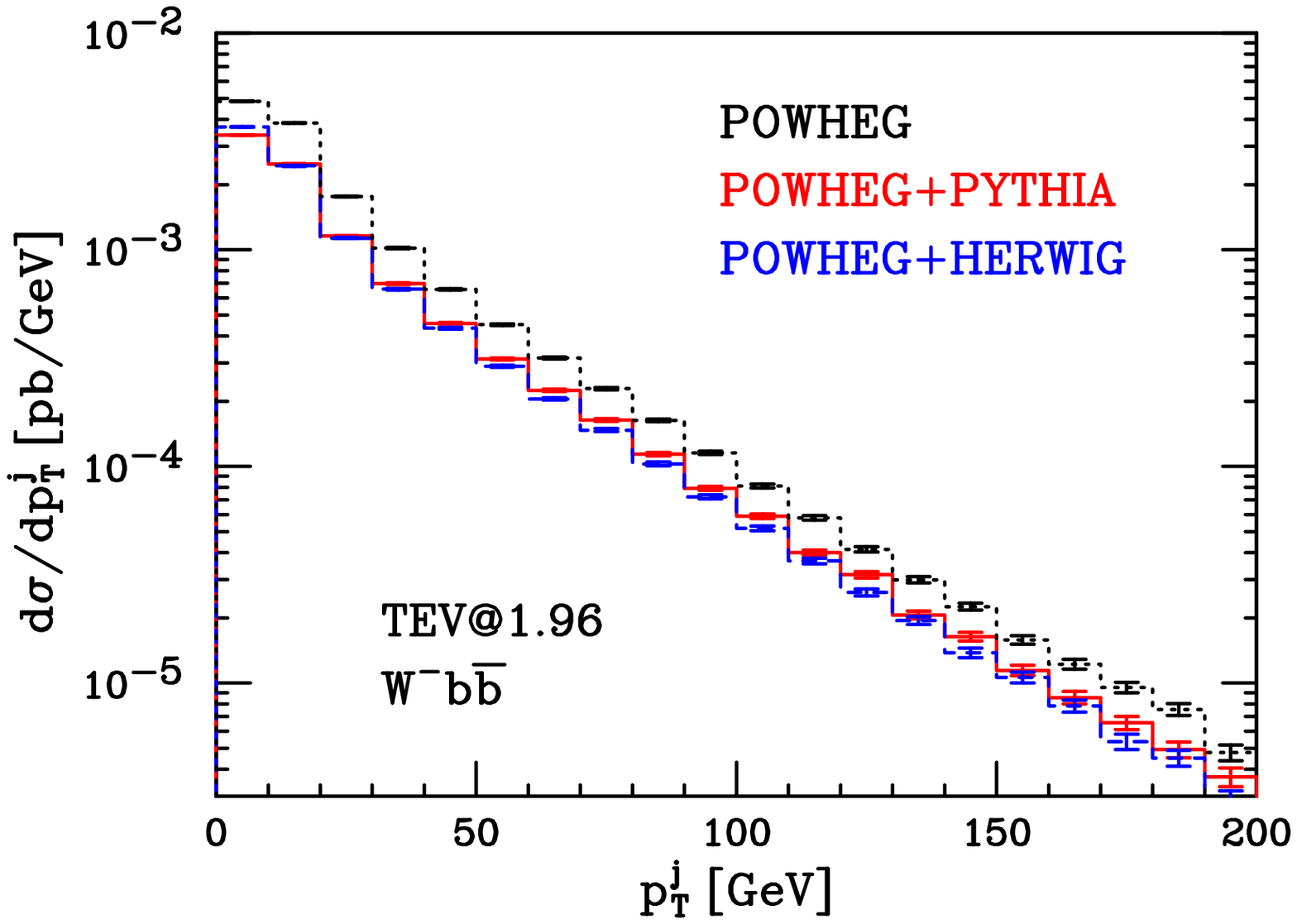,width=\wfig,height=\hfig}
\epsfig{file=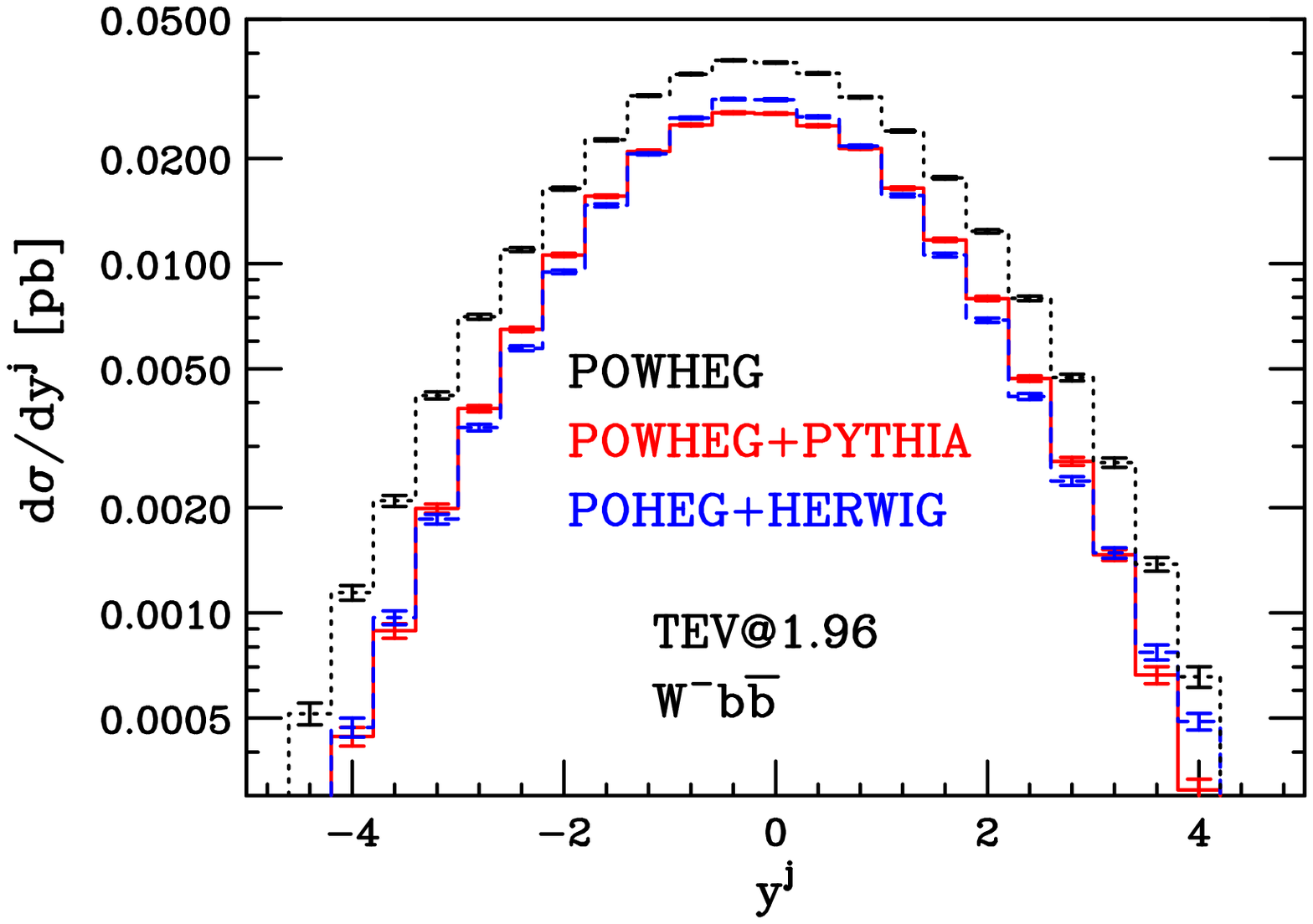,width=\wfig,height=\hfig}
\end{center}
\caption{\label{fig:TEV_PY_HW_2} Differential cross sections as a function of
  the transverse momentum, $p_T^j$, and the rapidity, $y^j$, of the hardest
  radiated non-$b$ jet, for $\Wmbbdec$ production at the Tevatron. The
  different curves represent the results for the \POWHEG{} hardest emission
  (dotted black), and for \POWHEG{} interfaced with either \PYTHIA{} (solid
  red) or \HERWIG{} (dashed blue).}
\end{figure}

\begin{figure}[htb]
\begin{center}
\epsfig{file=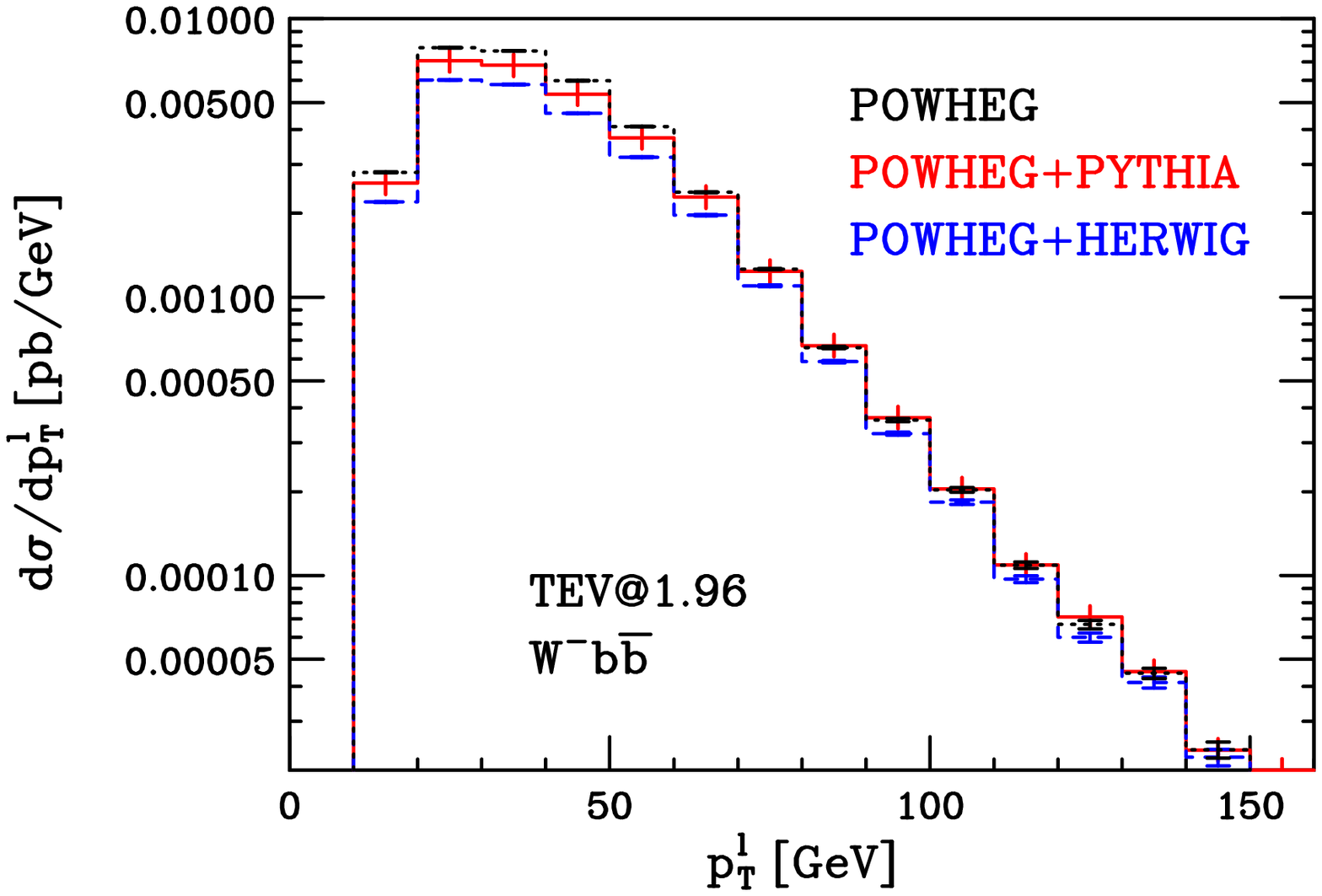,width=\wfig,height=\hfig}
\epsfig{file=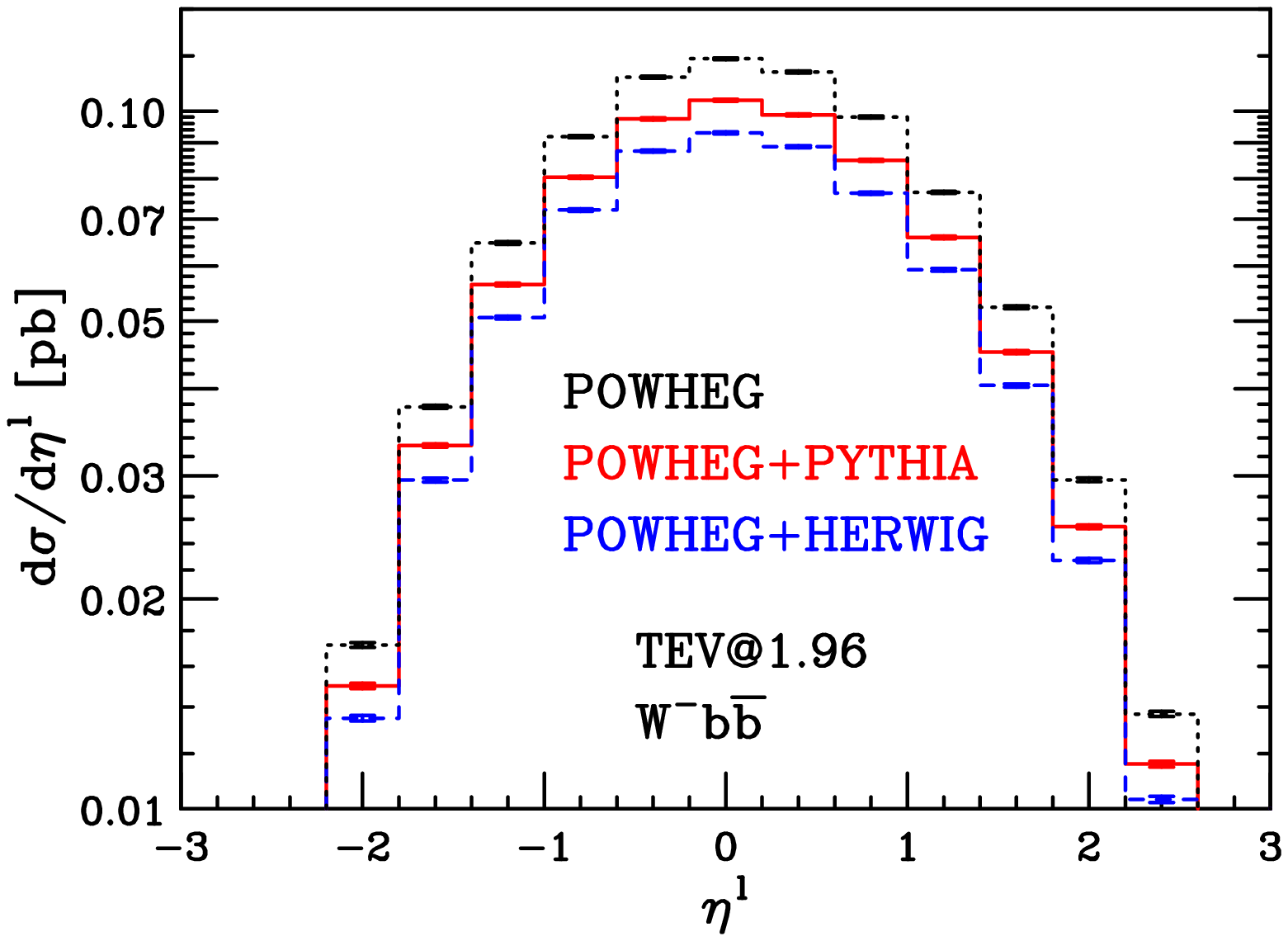,width=\wfig,height=\hfig}
\end{center}
\caption{\label{fig:TEV_PY_HW_3} Differential cross sections as a function of
  the transverse momentum, $p_T^l$, and the pseudorapidity, $\eta^l$, of the
  lepton for $\Wmbbdec$ production at the Tevatron. The different curves
  represent the results for the \POWHEG{} hardest emission (dotted black),
  and for \POWHEG{} interfaced with either \PYTHIA{} (solid red) or \HERWIG{}
  (dashed blue).}
\end{figure}

\begin{figure}[htb]
\begin{center}
\epsfig{file=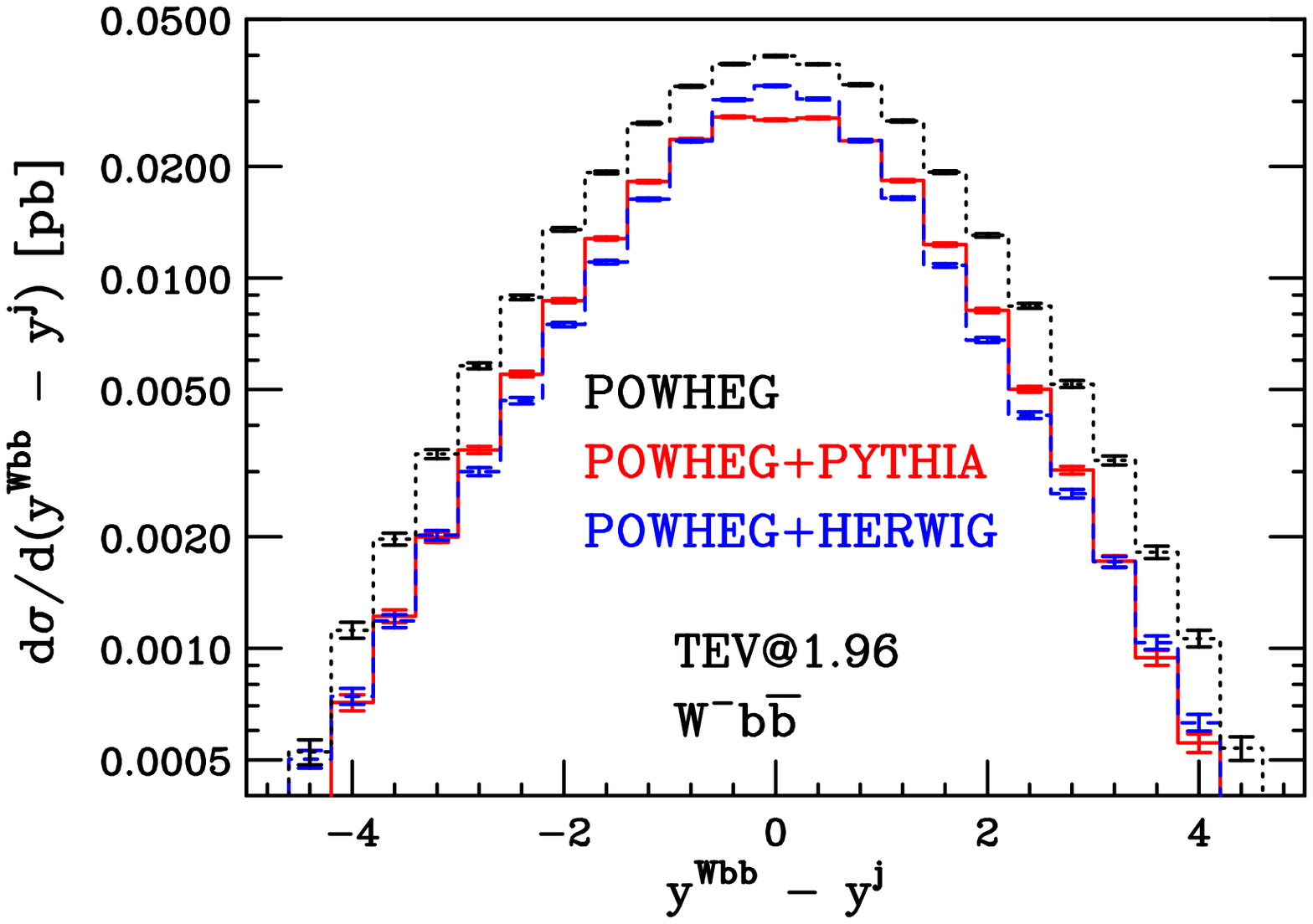,width=\wfig,height=\hfig}
\epsfig{file=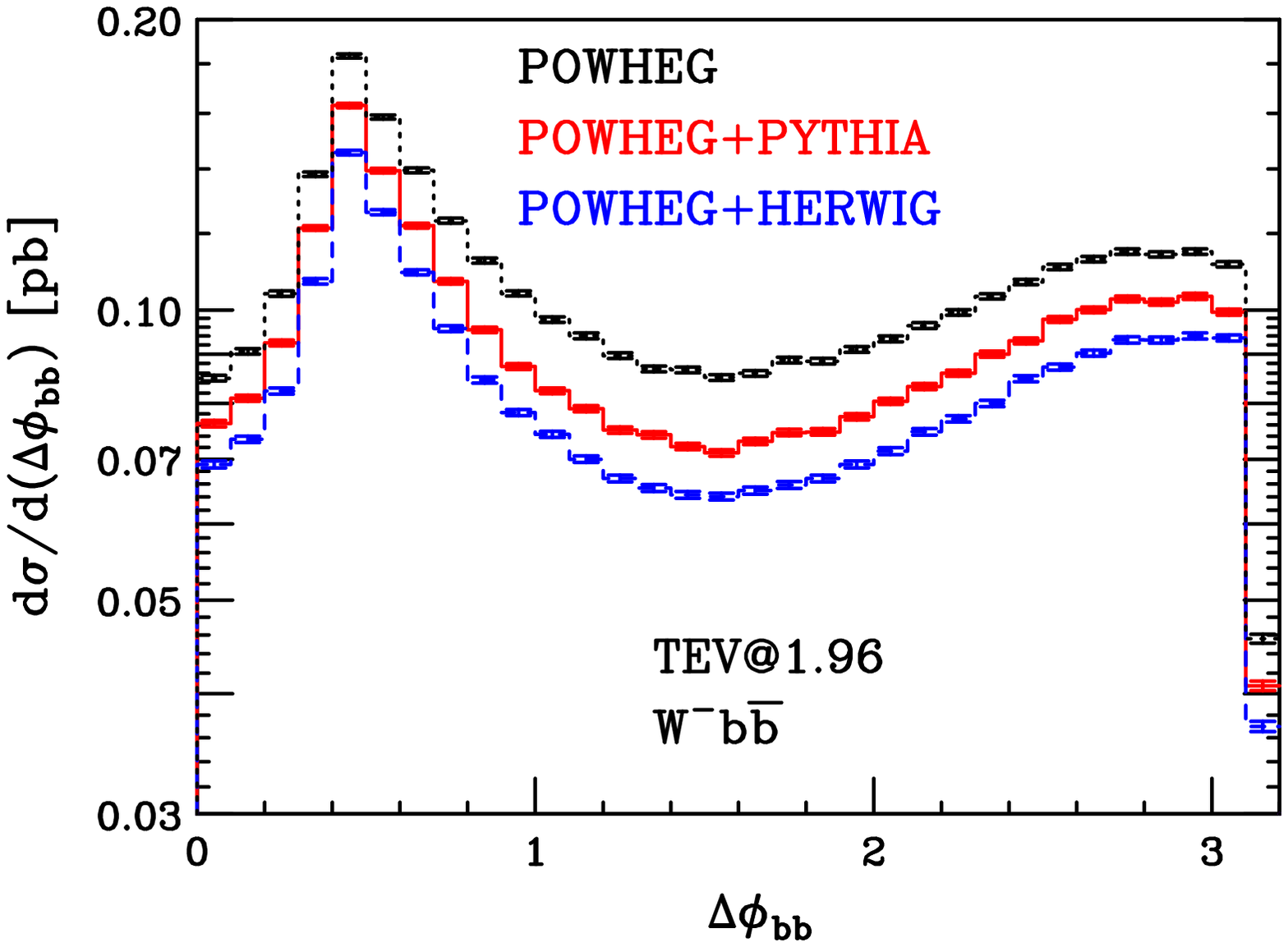,width=\wfig,height=\hfig}
\end{center}
\caption{\label{fig:TEV_PY_HW_4} Differential cross sections as a function of
  the rapidity difference between the $\Wbb$ system and the hardest
  radiated non-$b$ jet, $(y^{Wbb}-y^j)$, and the azimuthal angle
  difference between the two $b$ jets $\Delta\phi_{bb}$, for
  $\Wmbbdec$ production at the Tevatron. The different curves represent
  the results for the \POWHEG{} hardest emission (dotted black), and for
  \POWHEG{} interfaced with either \PYTHIA{} (solid red) or \HERWIG{} (dashed
  blue).}
\end{figure}

In fig.~\ref{fig:TEV_PY_HW_2}, we have plotted the differential cross
sections as a function of the transverse momentum of the hardest radiated
(non-$b$) jet, $\pt^j$, and its rapidity $y^j$, while in
fig.~\ref{fig:TEV_PY_HW_3} we have plotted the cross sections as a function
of the transverse momentum $\pt^l$ and pseudorapidity $\eta^l$ of the hardest
charged lepton. Finally, in fig.~\ref{fig:TEV_PY_HW_4}, we show the
differential cross sections as a function of $(y^{Wbb}-y^j)$, i.e.~the
difference between the rapidity of the $\Wbb$ system (the $W$ momentum is
taken from the showering program, since the neutrino goes undetected) and the
rapidity of the hardest jet, and as a function of the difference in the
azimuthal angles of the two $b$ jets, $\Delta\phi_{bb}$.

In all the plots, we can see that the different showers implemented by
\PYTHIA{} and \HERWIG{} give slightly different distributions, and we can
consider this difference as a theoretical error associated to showering
effects.
\begin{figure}[htb]
\begin{center}
\epsfig{file=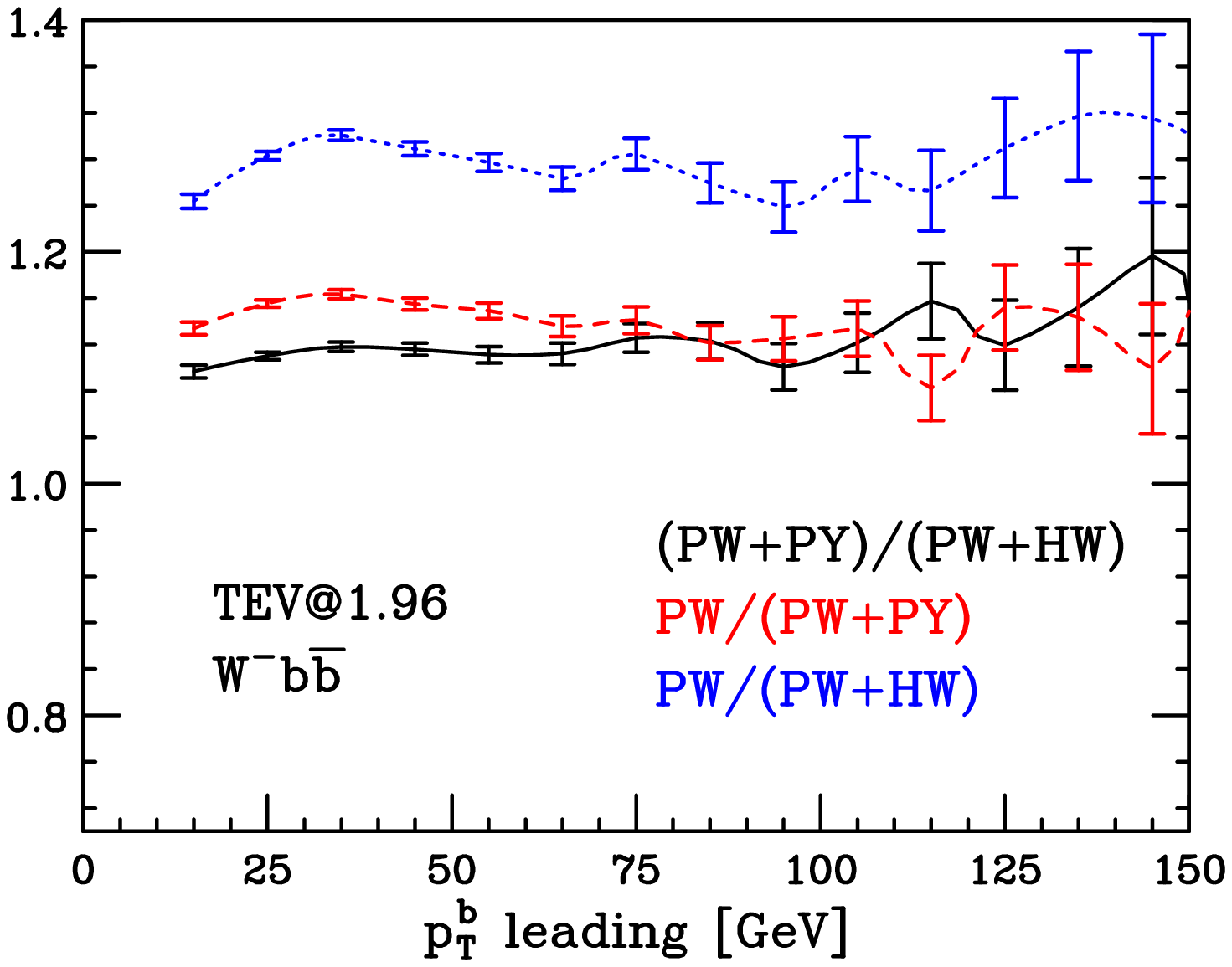,width=\wfig,height=\hfig}
\epsfig{file=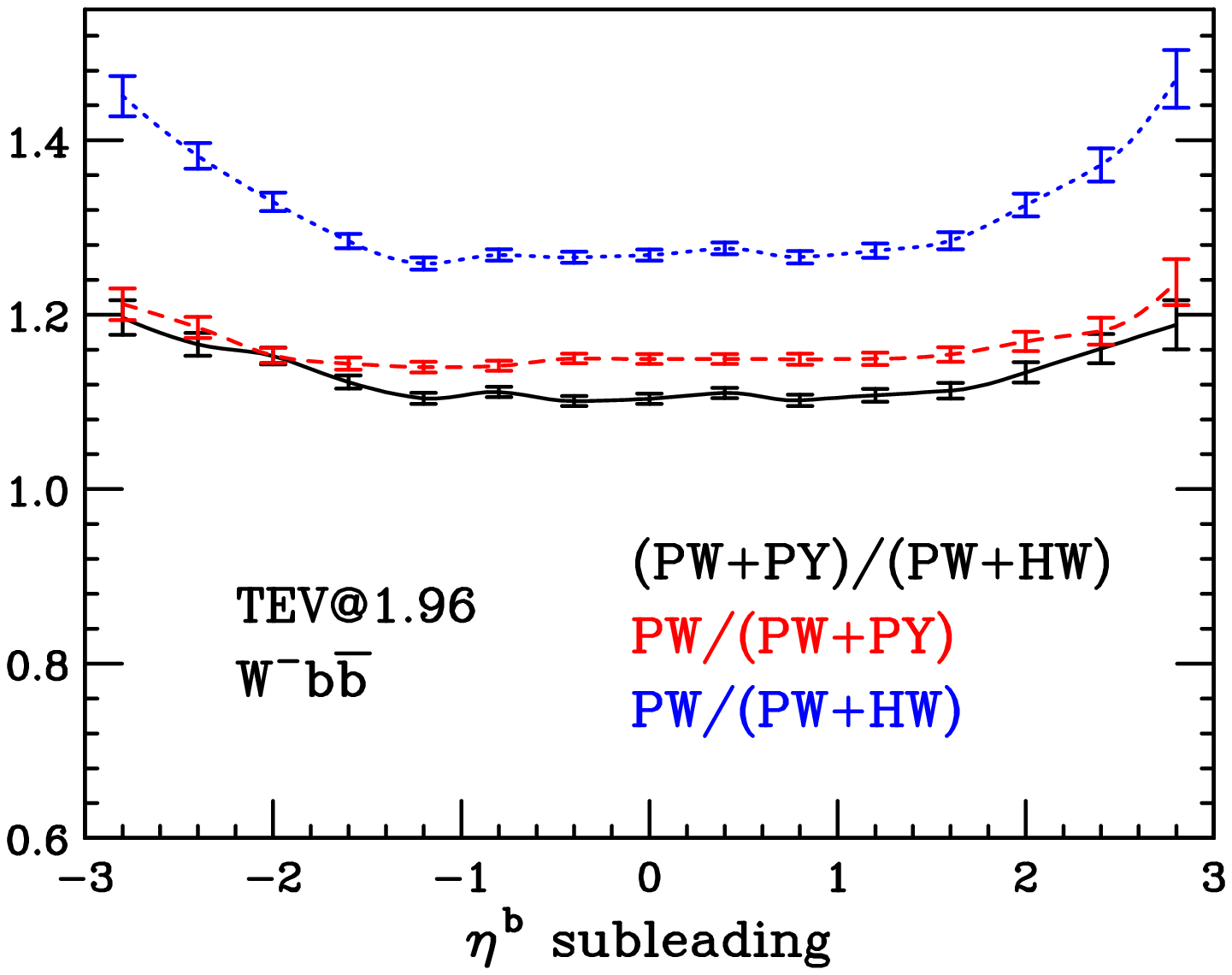,width=\wfig,height=\hfig}

\epsfig{file=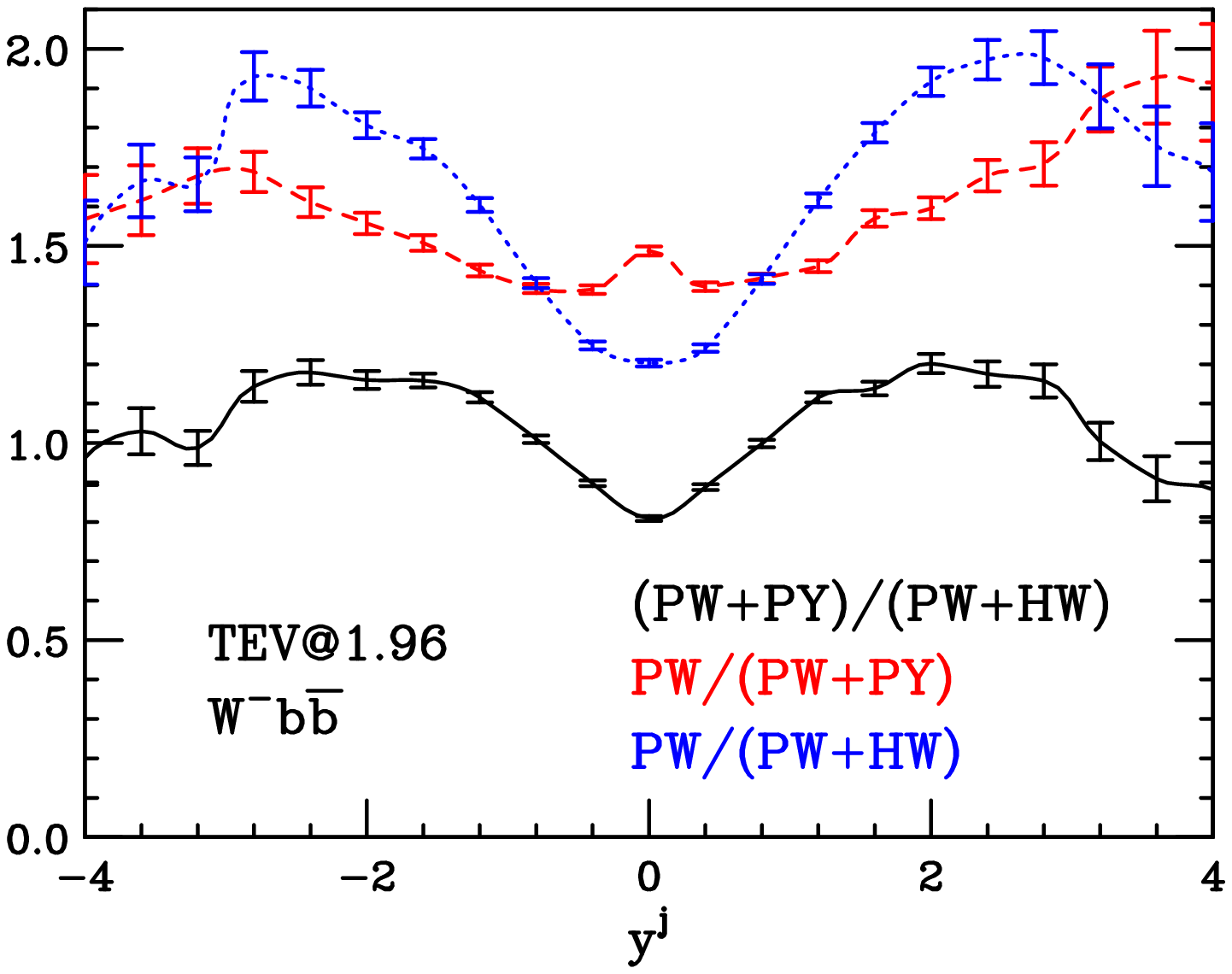,width=\wfig,height=\hfig}
\epsfig{file=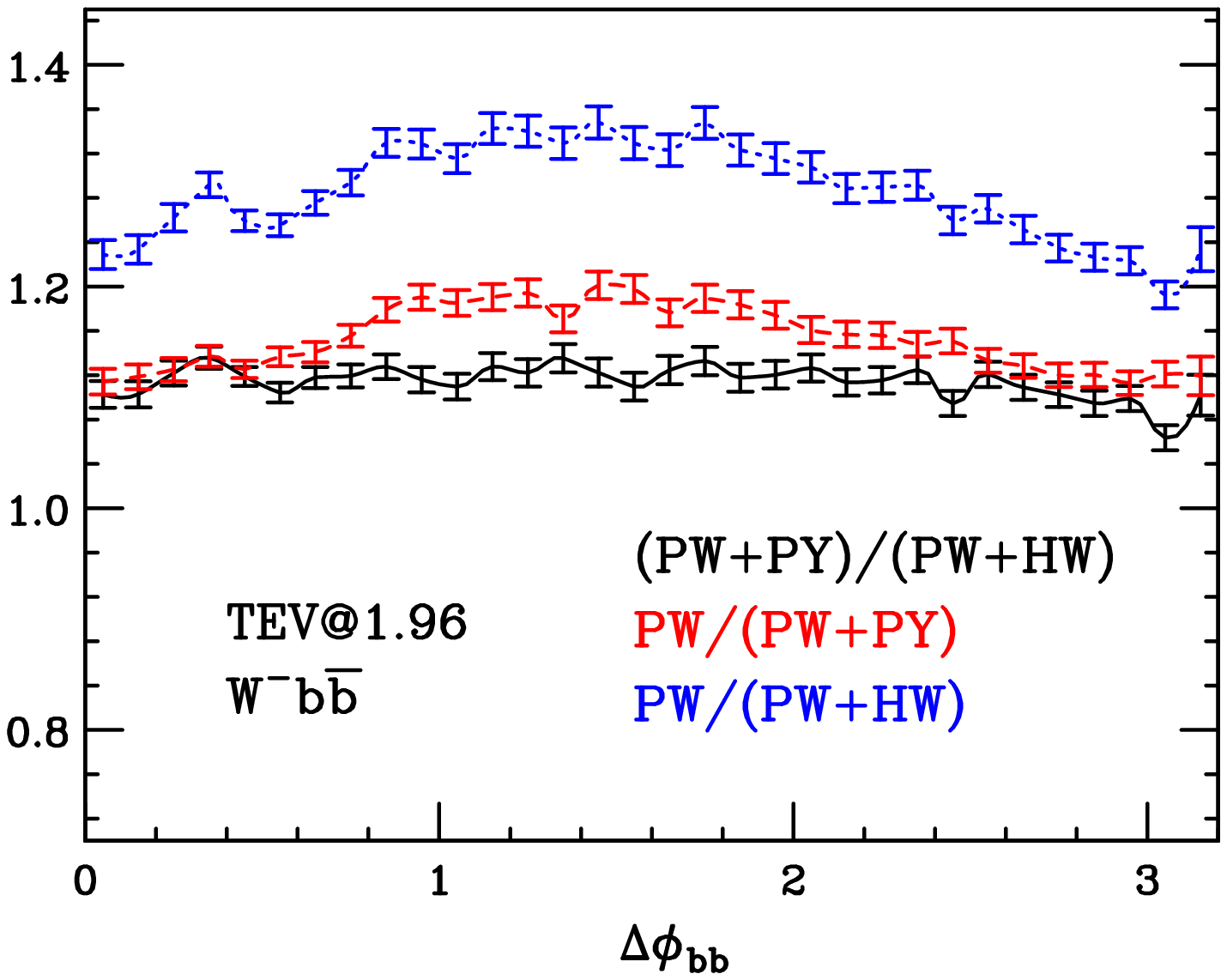,width=\wfig,height=\hfig}
\end{center}
\caption{\label{fig:TEV_PY_HW_ratio} Ratios of the differential cross
  sections for $\Wmbbdec$ production at the Tevatron:
  (\POWHEG{}+\PYTHIA{})/ (\POWHEG{}+\HERWIG{}) in solid black lines,
  \POWHEG{}/(\POWHEG{}+\PYTHIA{}) in dashed red lines and
  \POWHEG{}/(\POWHEG{}+\HERWIG{}) in dotted blue lines. Starting from the
  upper left corner and moving clockwise we show the ratio of the
  differential cross sections as function of the transverse momentum of the
  hardest $b$ jet, $p_T^b$ leading, the pseudorapidity of the second hardest
  $b$ jet, $\eta^b$ subleading, the azimuthal angular difference between the
  two $b$ jets, $\Delta\phi_{bb}$, and the rapidity of the hardest radiated
  non $b$ jet, $y^j$.}
\end{figure}
The trend of the distributions is the same in all the plots: the differential
cross sections from \POWHEG{} followed by the shower done by \PYTHIA{}
(\POWHEG+\PYTHIA) are slightly larger than the corresponding ones showered by
\HERWIG{} (\POWHEG+\HERWIG). A consequence of this fact is that the cross
sections, after the cuts of eq.~(\ref{eq:cuts}), are given by
\begin{equation}
\sigma_{\POWHEG} = 0.335~\pb,\qquad
\sigma_{\POWHEG+\PYTHIA} = 0.291~\pb,\qquad
\sigma_{\POWHEG+\HERWIG} = 0.262~\pb.
\end{equation} 
In order to evidentiate the differences between the \POWHEG{} ({\tt PW}), the
\POWHEG+\PYTHIA{} ({\tt PW+PY}) and the \POWHEG+\HERWIG{} ({\tt PW+HW})
results, in fig.~\ref{fig:TEV_PY_HW_ratio} we have plotted their ratio for a
sample of distributions.  While the ratios of the \POWHEG{} hardest emission
cross sections over the \POWHEG{} showered results ({\tt PW/(PW+PY)} for the
shower done by \PYTHIA{}, in dashed red lines, and {\tt PW/(PW+HW)} for the
shower done by \HERWIG{}, in dotted blue lines) are just an indication of the
effect of the completion of the shower on the \POWHEG{} hardest-emission
events, the ratios {\tt (PW+PY)/(PW+HW)} carry information on the effects of
the two different showering algorithms implemented in \PYTHIA{} and
\HERWIG{}.
As can be inferred from the {\tt (PW+PY)/(PW+HW)} ratios in the figures,
these effects amount to differences of the order of 10--20\%, and this can be
taken as the theoretical errors connected with using different showering
programs. In general, the ratios are almost flat. The only distribution that
shows some phase-space dependence is the rapidity of the hardest jet, in the
lower left plot of fig.~\ref{fig:TEV_PY_HW_ratio}, corresponding to the
ratios of the curves in the right-hand-side plot in
fig.~\ref{fig:TEV_PY_HW_2}. Here, {\tt PW+HW} jets tend to be more central in
rapidity then the {\tt PW+PY} ones. We expect this same behaviour to be
present in fig.~\ref{fig:TEV_PY_HW_4} as well, since we are plotting the
differential distribution as a function of $(y^{Wbb}-y^j)$, and the rapidity
of the jet enters directly in this quantity.

\subsection{LHC results}
\label{sec:phenomenolgy_lhc}

\begin{figure}[htb]
\begin{center}
\epsfig{file=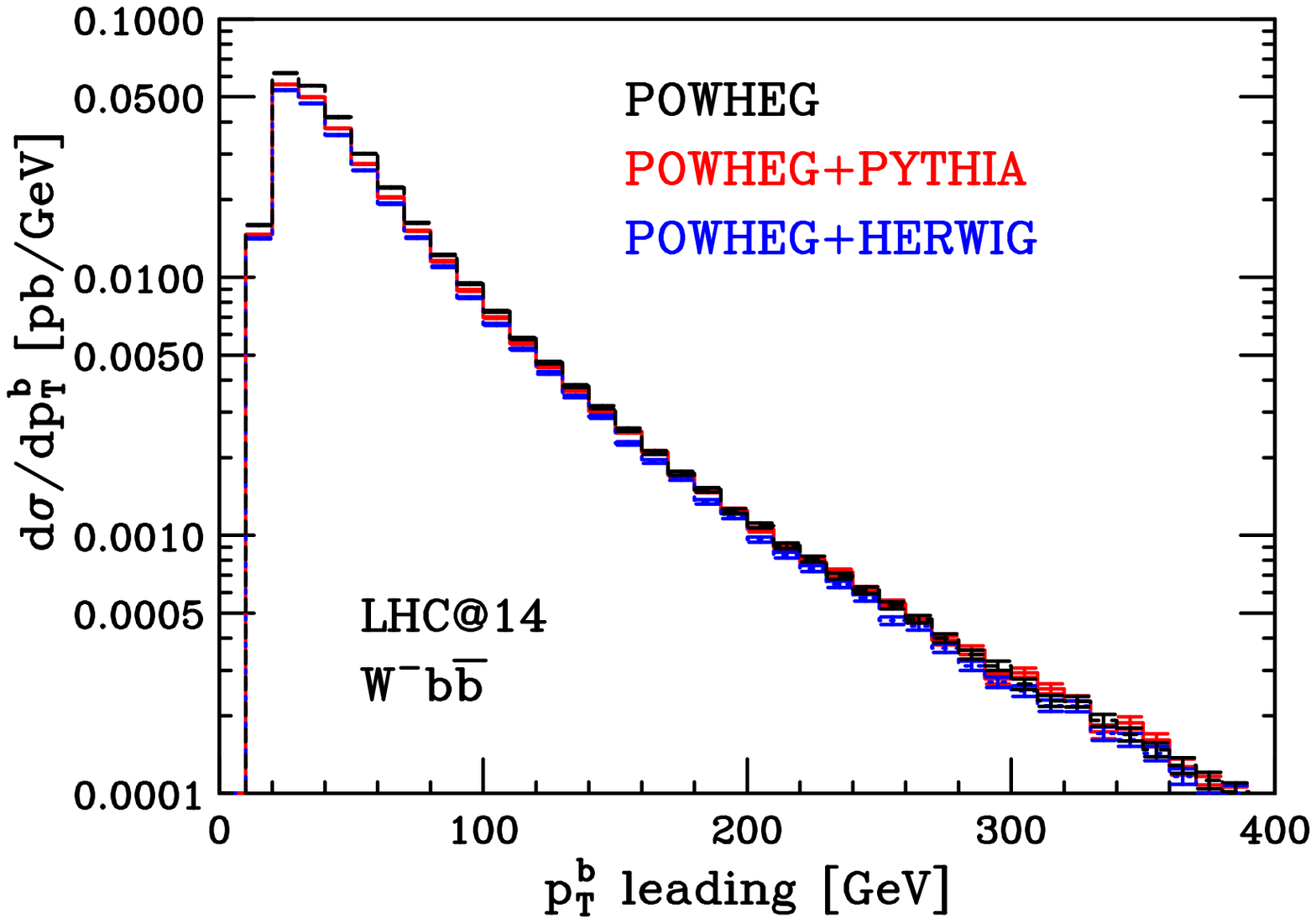,width=\wfig,height=\hfig}
\epsfig{file=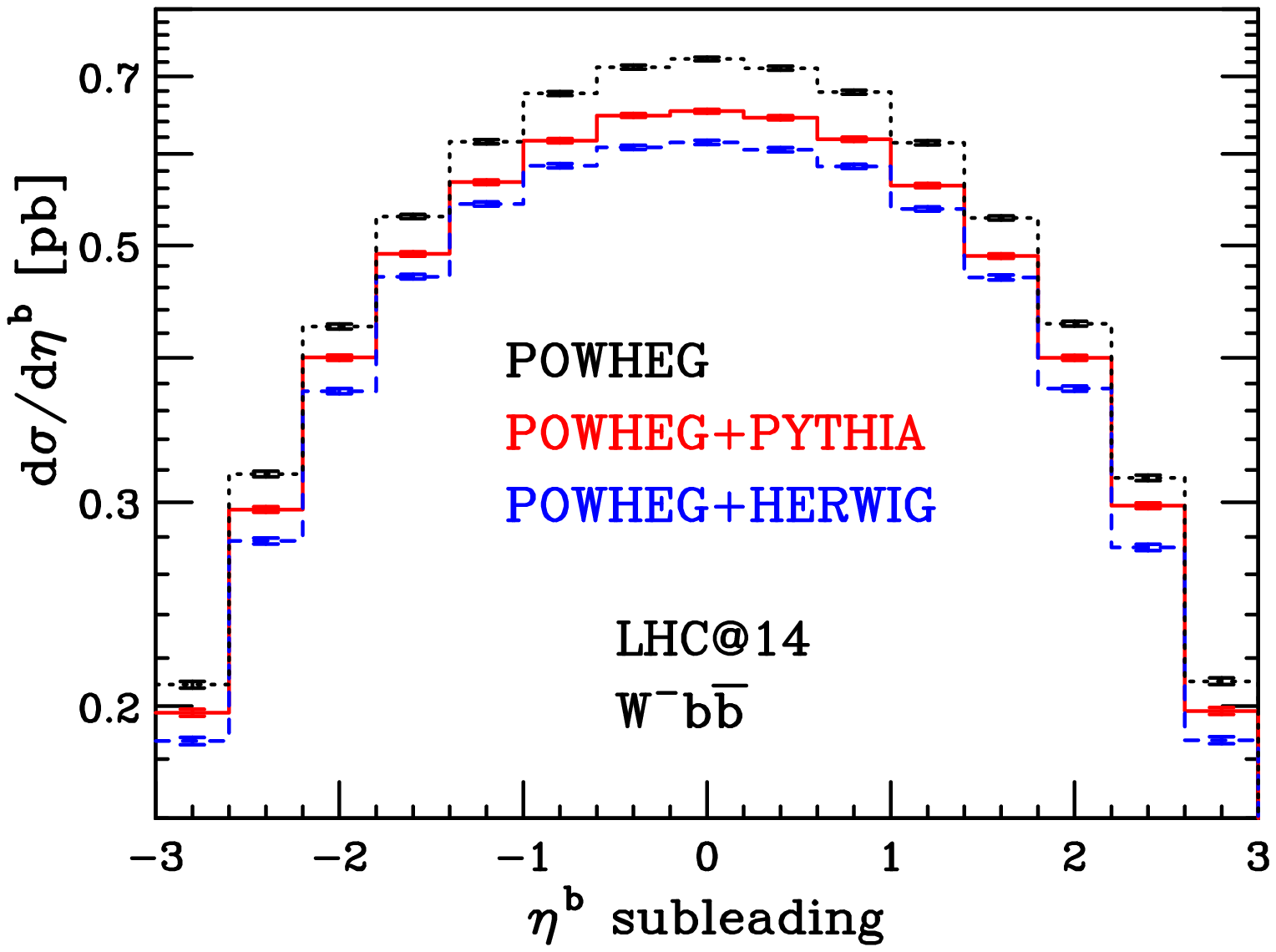,width=\wfig,height=\hfig}

\epsfig{file=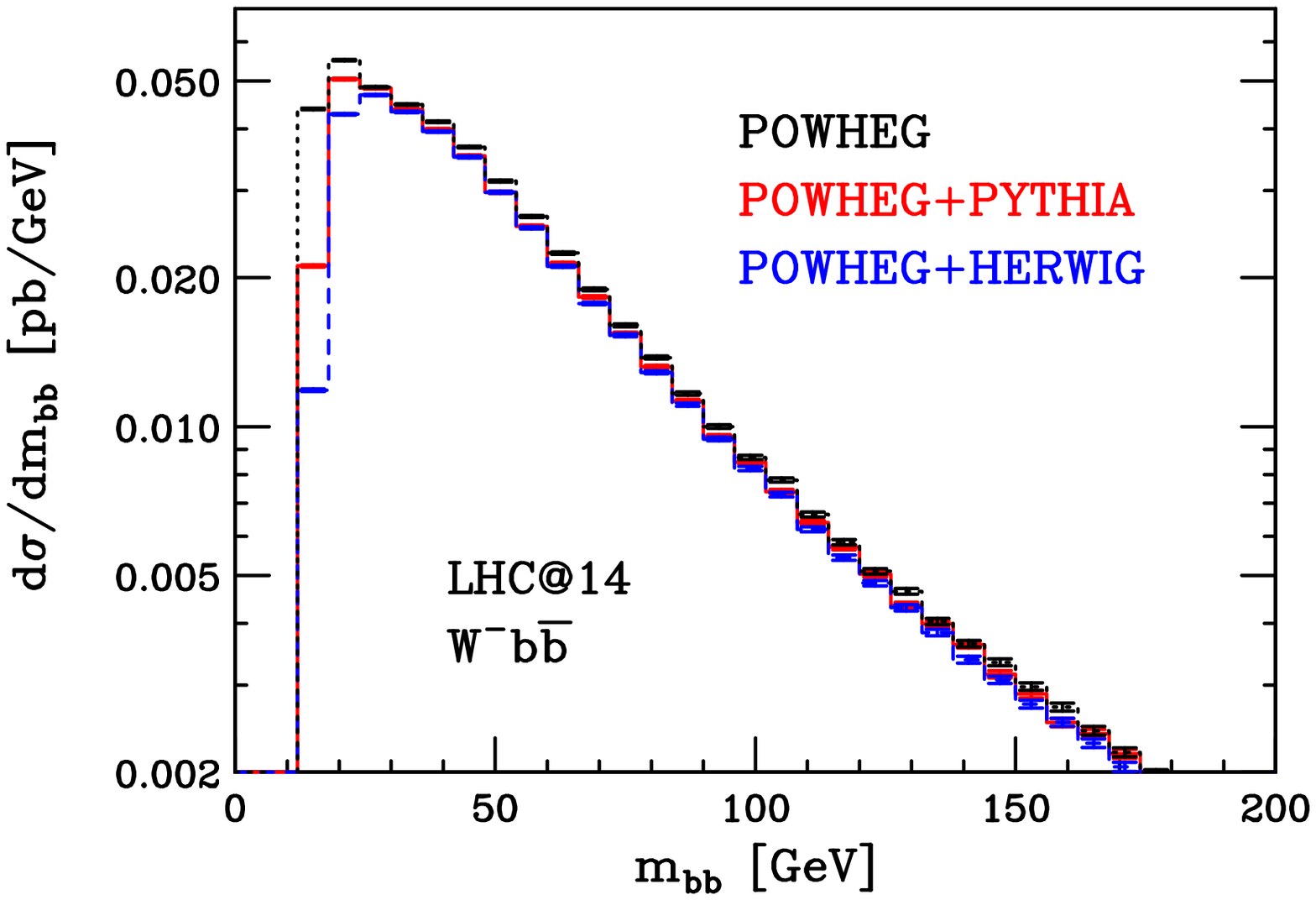,width=\wfig,height=\hfig}
\epsfig{file=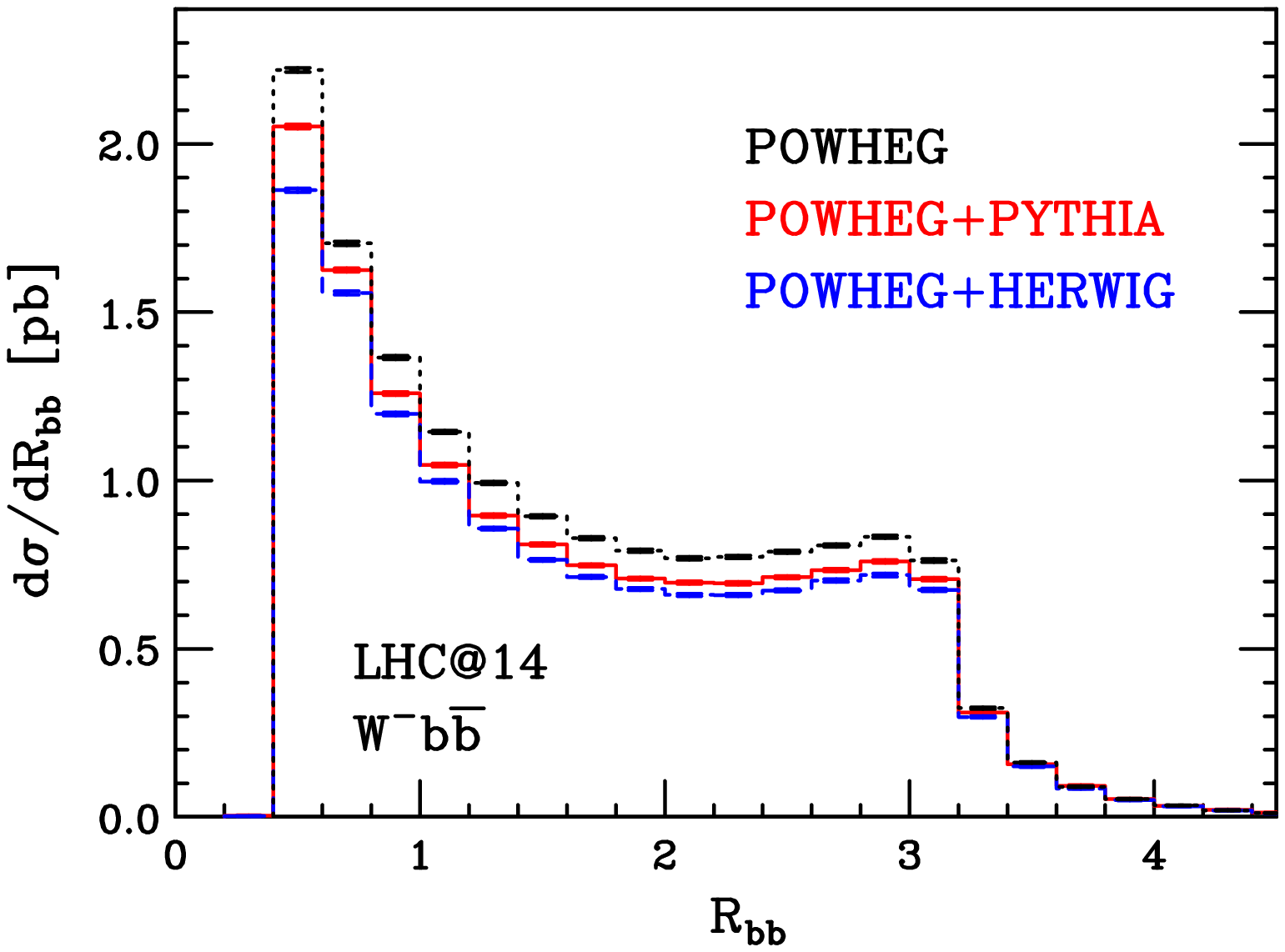,width=\wfig,height=\hfig}
\end{center}
\caption{\label{fig:LHC_PY_HW_1} Differential cross section as a
  function of the transverse momentum of the hardest $b$ jet, $p_T^b$
  leading, the pseudorapidity of the second hardest $b$ jet, $\eta^b$
  subleading, the invariant mass of the leading and subleading $b$
  jets $m_{bb}$, and their angular distance $R_{bb}$, for
  $\Wmbbdec$ production at the LHC with $\sqrt{s}=14$~TeV. The
  different curves represent the results of \POWHEG{} hardest emission
  (dotted black), and of \POWHEG{} interfaced with either \PYTHIA{}
  (solid red) or \HERWIG{} (dashed blue).}
\end{figure}

\begin{figure}[htb]
\begin{center}
\epsfig{file=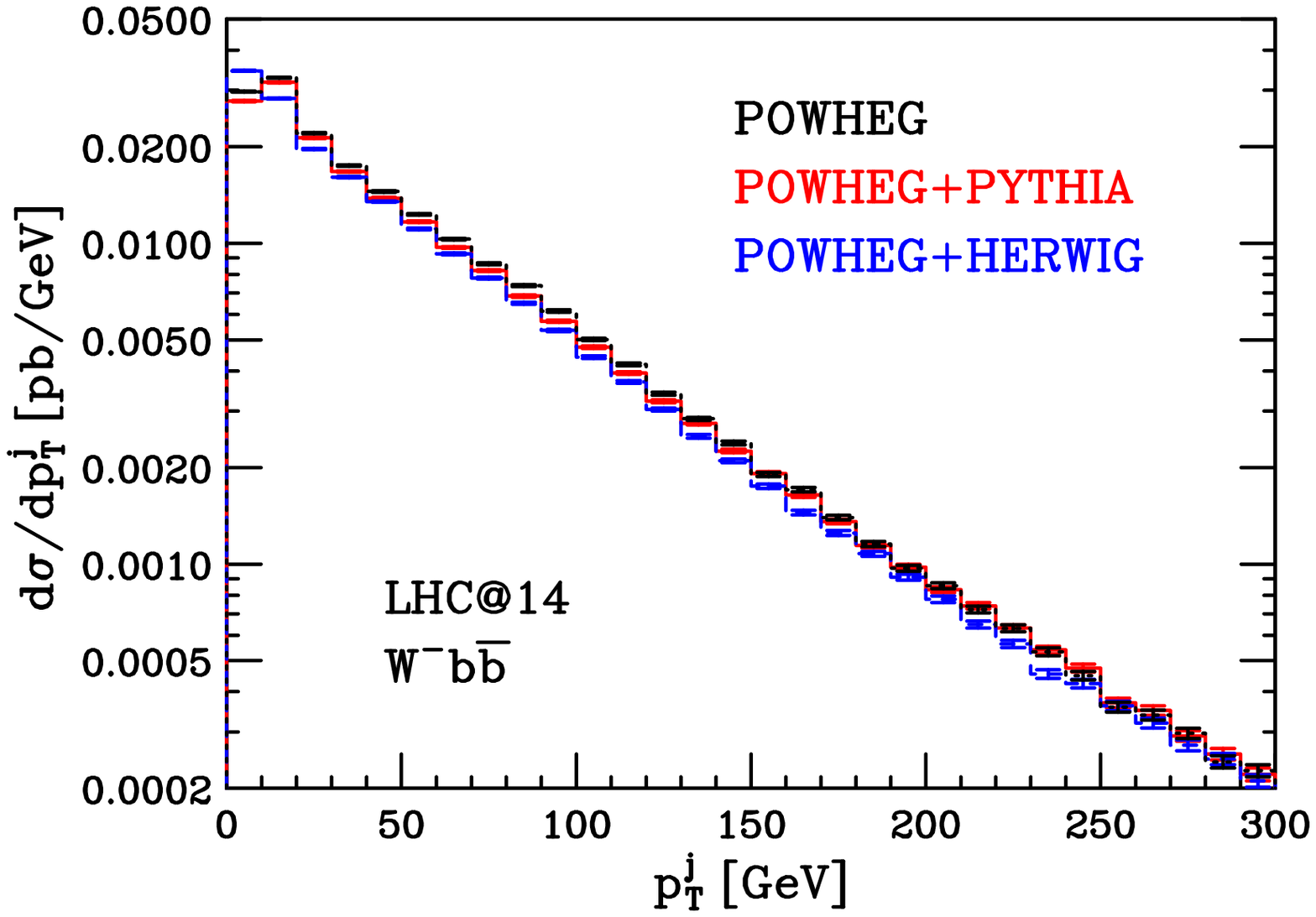,width=\wfig,height=\hfig}
\epsfig{file=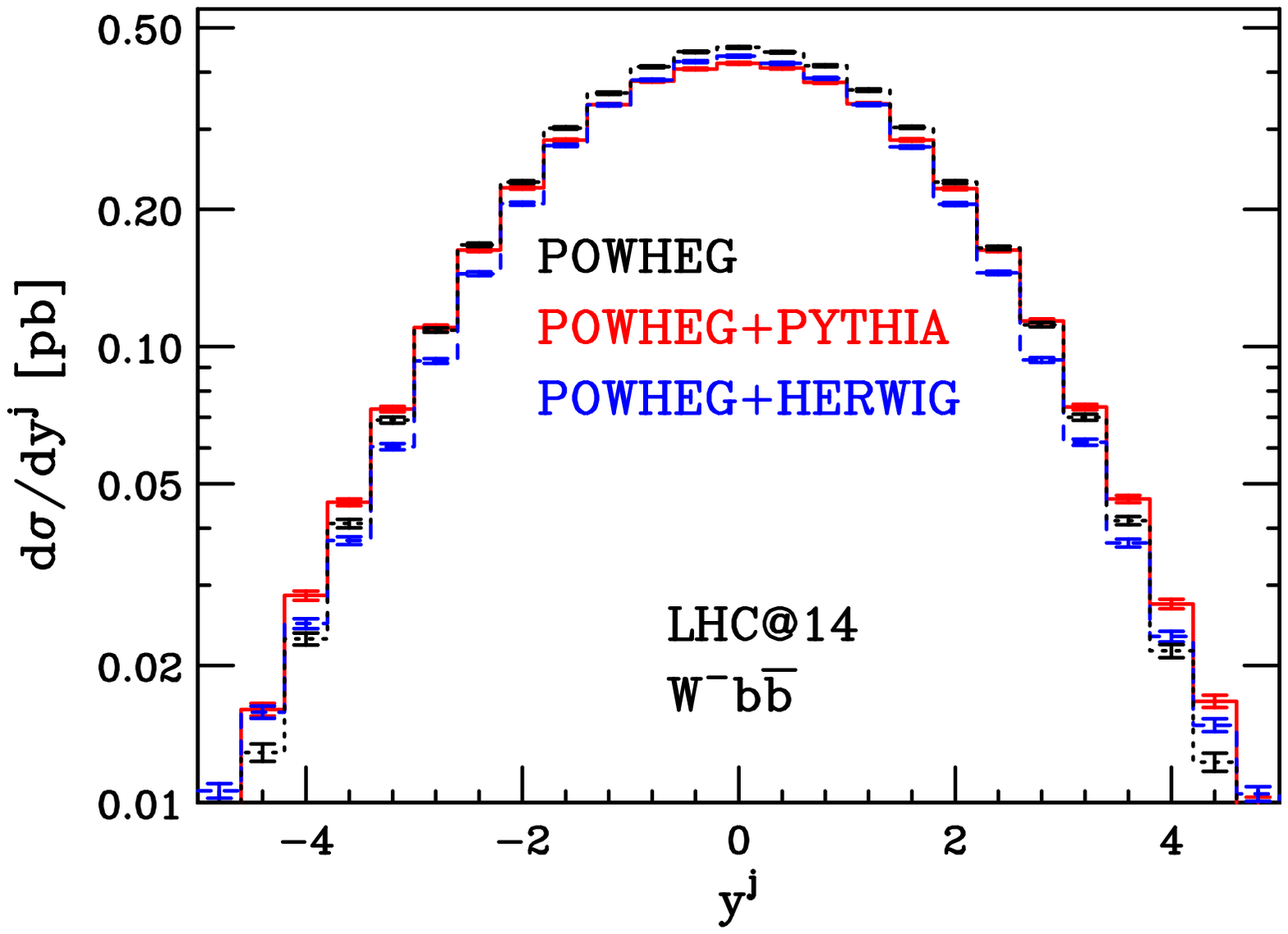,width=\wfig,height=\hfig}
\end{center}
\caption{\label{fig:LHC_PY_HW_2} Differential cross section as a
  function of the transverse momentum, $p_T^j$, and the rapidity $y^j$
  of the hardest radiated non-$b$ jet, for $\Wmbbdec$ production at
  the LHC with $\sqrt{s}=14$~TeV. The different curves represent the
  results of \POWHEG{} hardest emission (dotted black), and of
  \POWHEG{} interfaced with either \PYTHIA{} (solid red) or \HERWIG{}
  (dashed blue).}
\end{figure}

\begin{figure}[htb]
\begin{center}
\epsfig{file=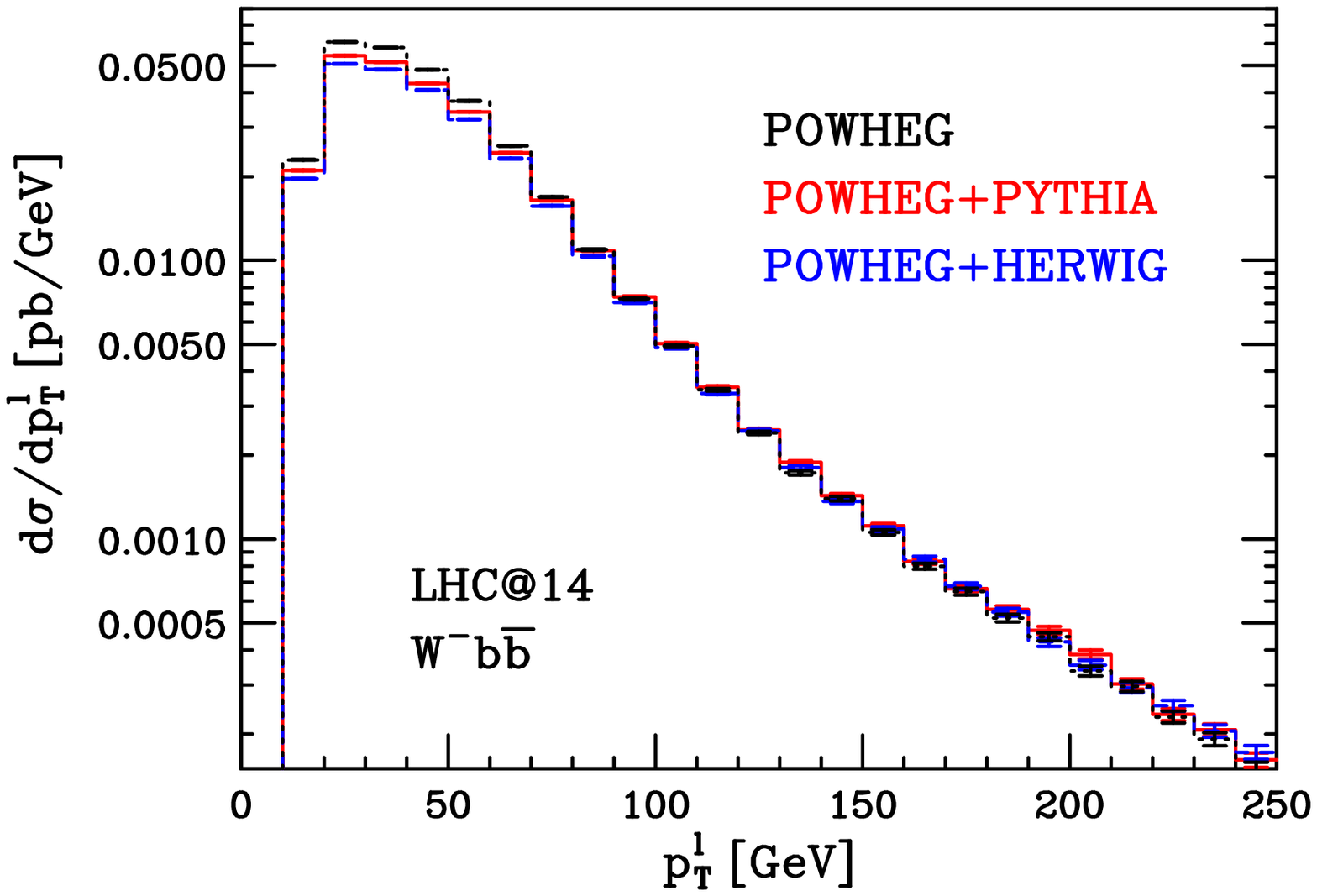,width=\wfig,height=\hfig}
\epsfig{file=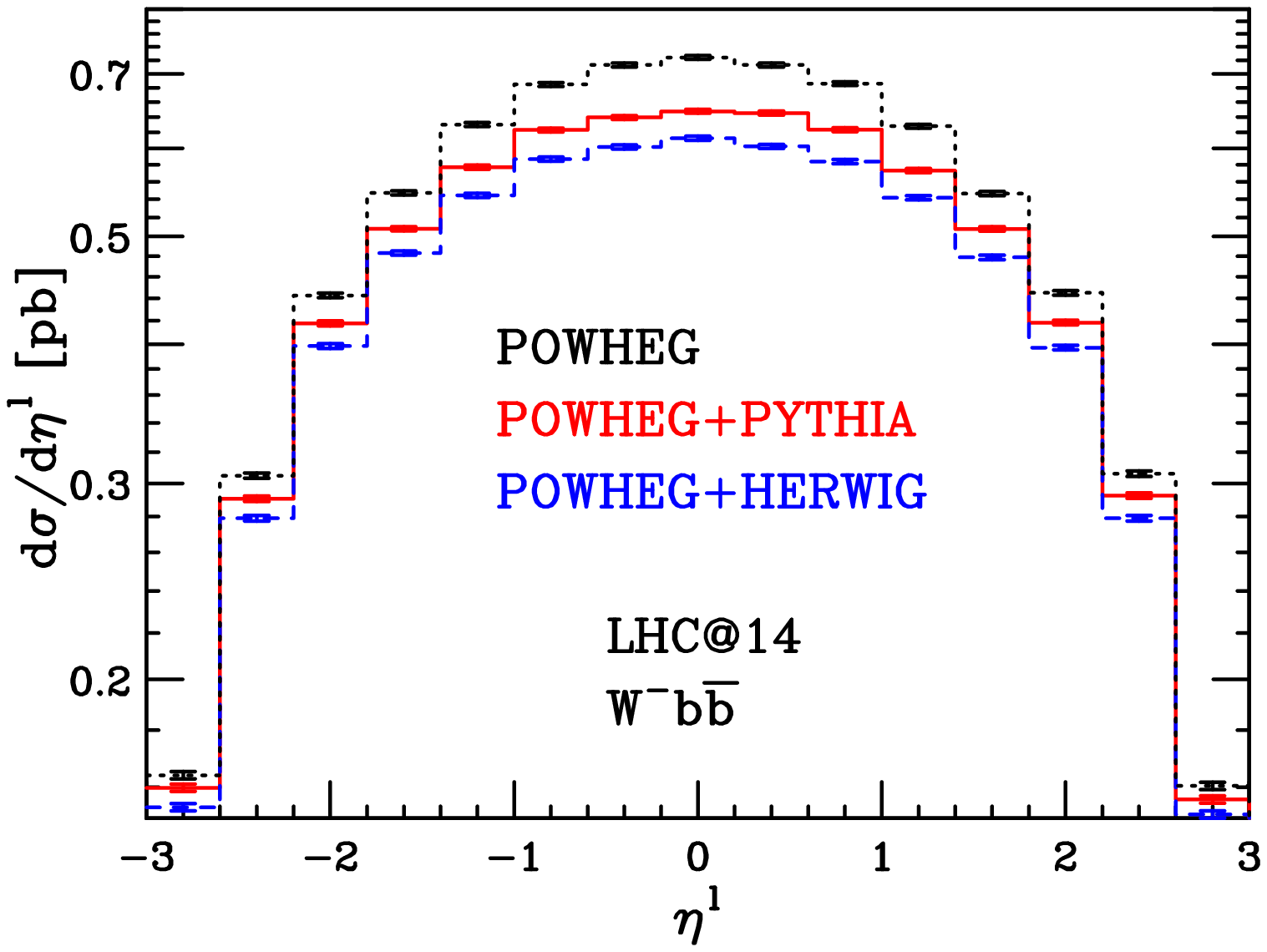,width=\wfig,height=\hfig}
\end{center}
\caption{\label{fig:LHC_PY_HW_3} Differential cross section as a function of
  the transverse momentum, $p_T^l$, and the pseudorapidity $\eta^l$ of the
  lepton for $\Wmbbdec$ production at the LHC with $\sqrt{s}=14$~TeV. The
  different curves represent the results of \POWHEG{} hardest emission
  (dotted black), and of \POWHEG{} interfaced with either \PYTHIA{} (solid
  red) or \HERWIG{} (dashed blue).}
\end{figure}

\begin{figure}[htb]
\begin{center}
\epsfig{file=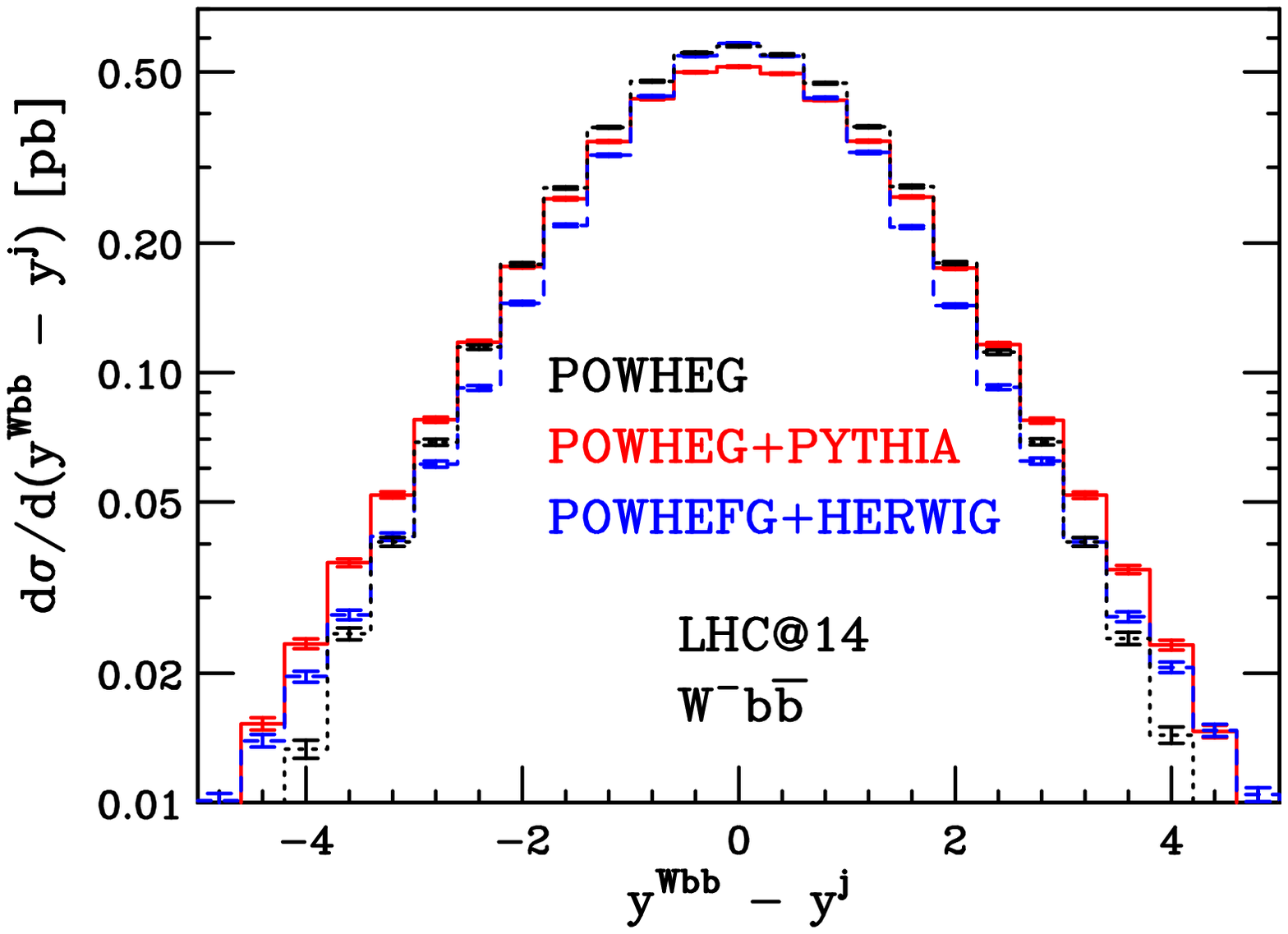,width=\wfig,height=\hfig}
\epsfig{file=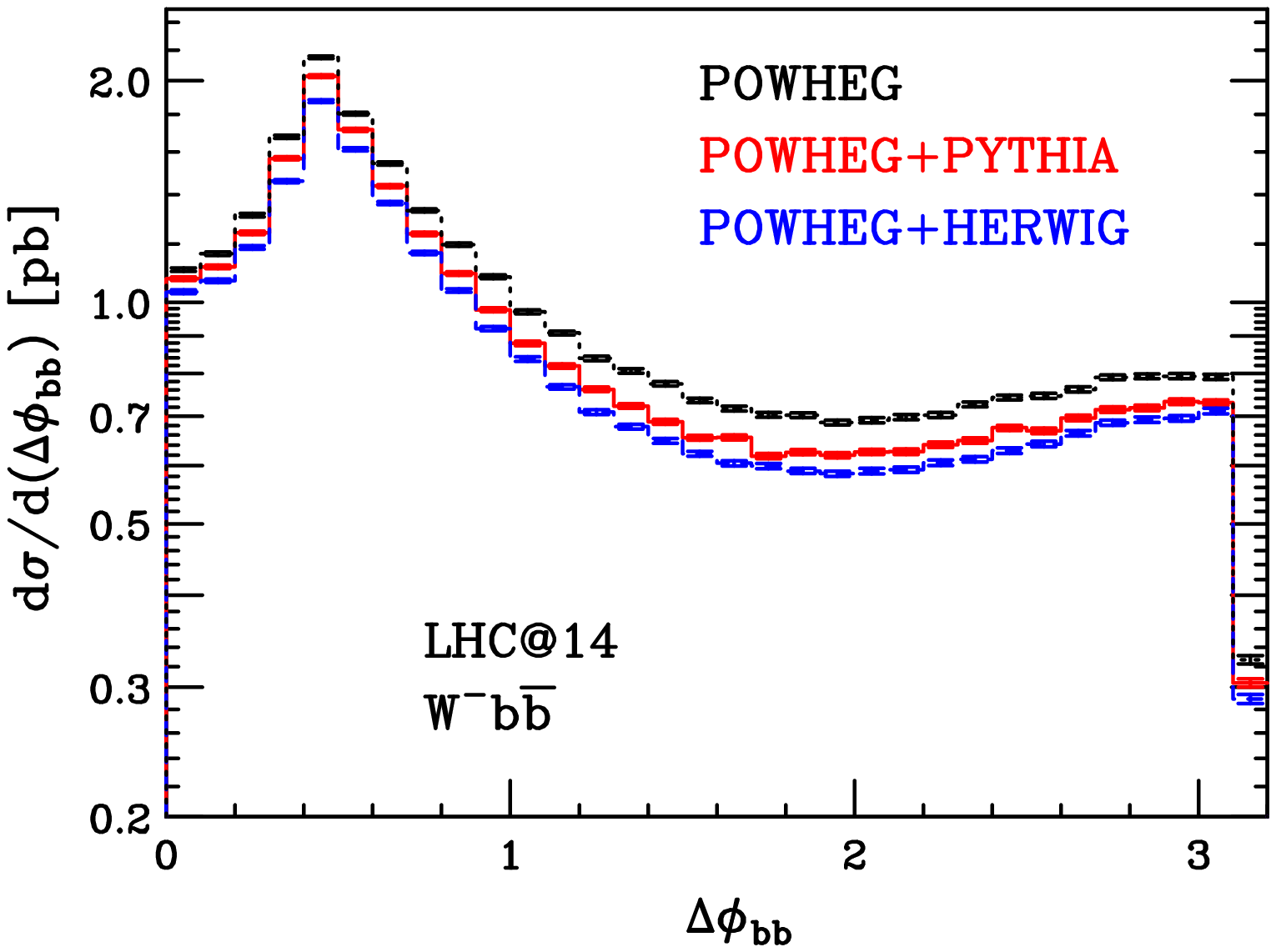,width=\wfig,height=\hfig}
\end{center}
\caption{\label{fig:LHC_PY_HW_4} Differential cross section as a
  function of the rapidity difference between the $\Wbb$ system
  and the hardest radiated non-$b$ jet, $(y^{Wbb}-y^j)$, and the
  azimuthal angle difference between the two $b$ jets
  $\Delta\phi_{bb}$, for $\Wmbbdec$ production at the LHC
  with $\sqrt{s}=14$~TeV. The different curves represent the results
  of \POWHEG{} hardest emission (dotted black), and of \POWHEG{}
  interfaced with either \PYTHIA{} (solid red) or \HERWIG{} (dashed
  blue).}
\end{figure}

In figs.~\ref{fig:LHC_PY_HW_1}, \ref{fig:LHC_PY_HW_2}, \ref{fig:LHC_PY_HW_3}
and~\ref{fig:LHC_PY_HW_4}, we have plotted the same differential cross
sections we have studied for the Tevatron, this time for the LHC with
$\sqrt{s}=14$~TeV.  The behaviour of all the distributions is the same as for
the Tevatron, and the corresponding cross sections after cuts are given by
\begin{equation}
\sigma_{\POWHEG} = 3.08~\pb,\qquad
\sigma_{\POWHEG+\PYTHIA} = 2.83~\pb,\qquad
\sigma_{\POWHEG+\HERWIG} = 2.68~\pb.
\end{equation}

\begin{figure}[htb]
\begin{center}
\epsfig{file=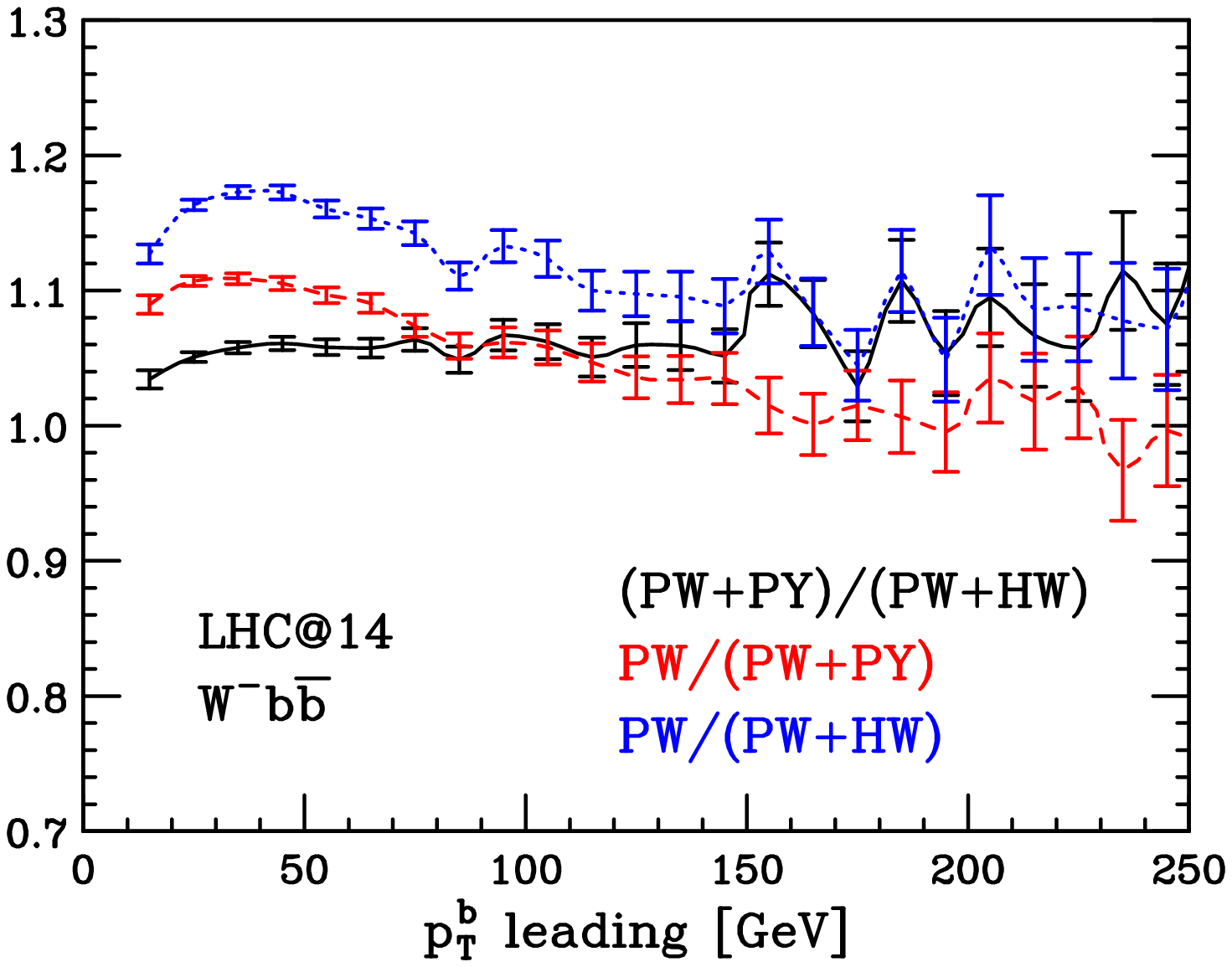,width=\wfig,height=\hfig}
\epsfig{file=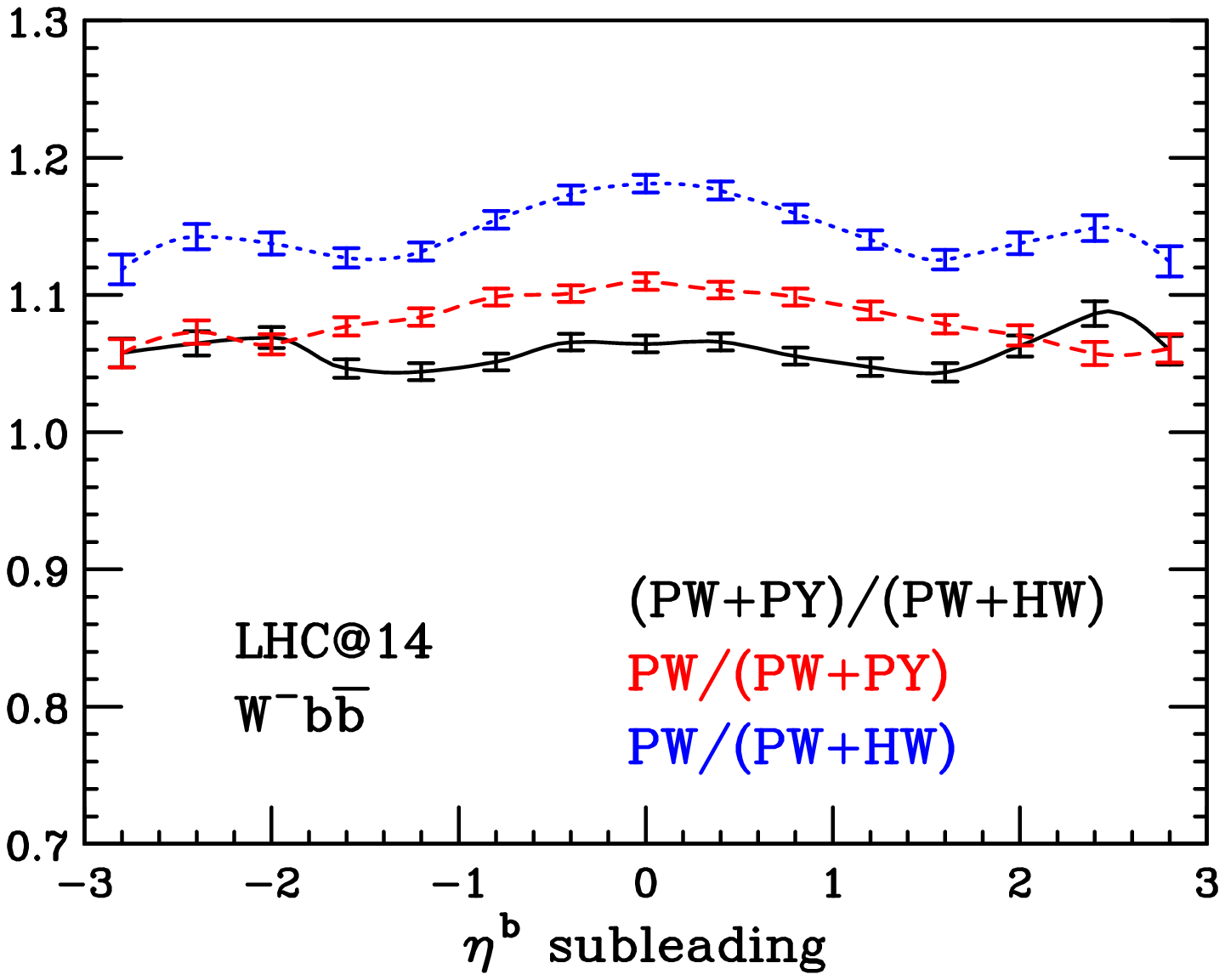,width=\wfig,height=\hfig}

\epsfig{file=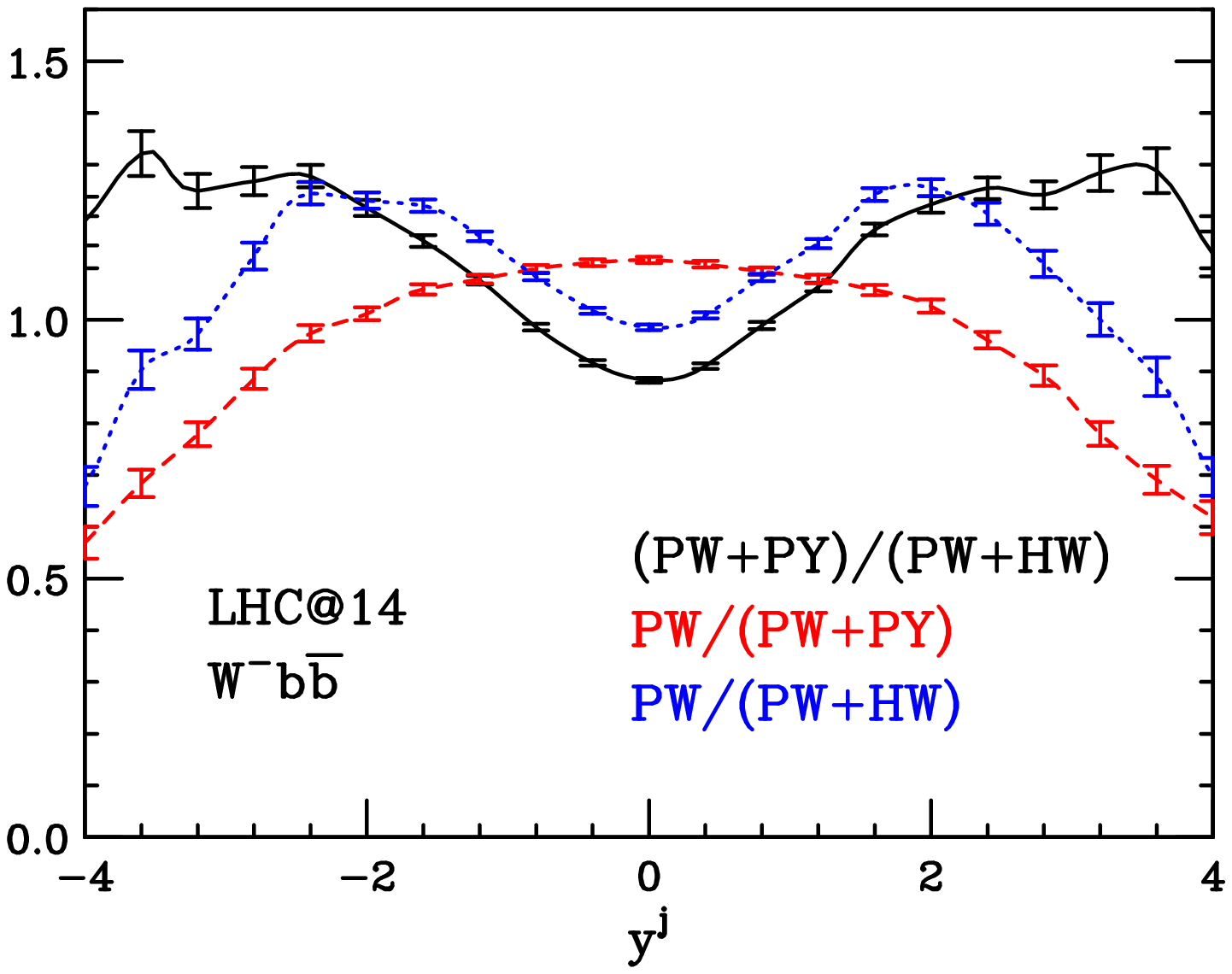,width=\wfig,height=\hfig}
\epsfig{file=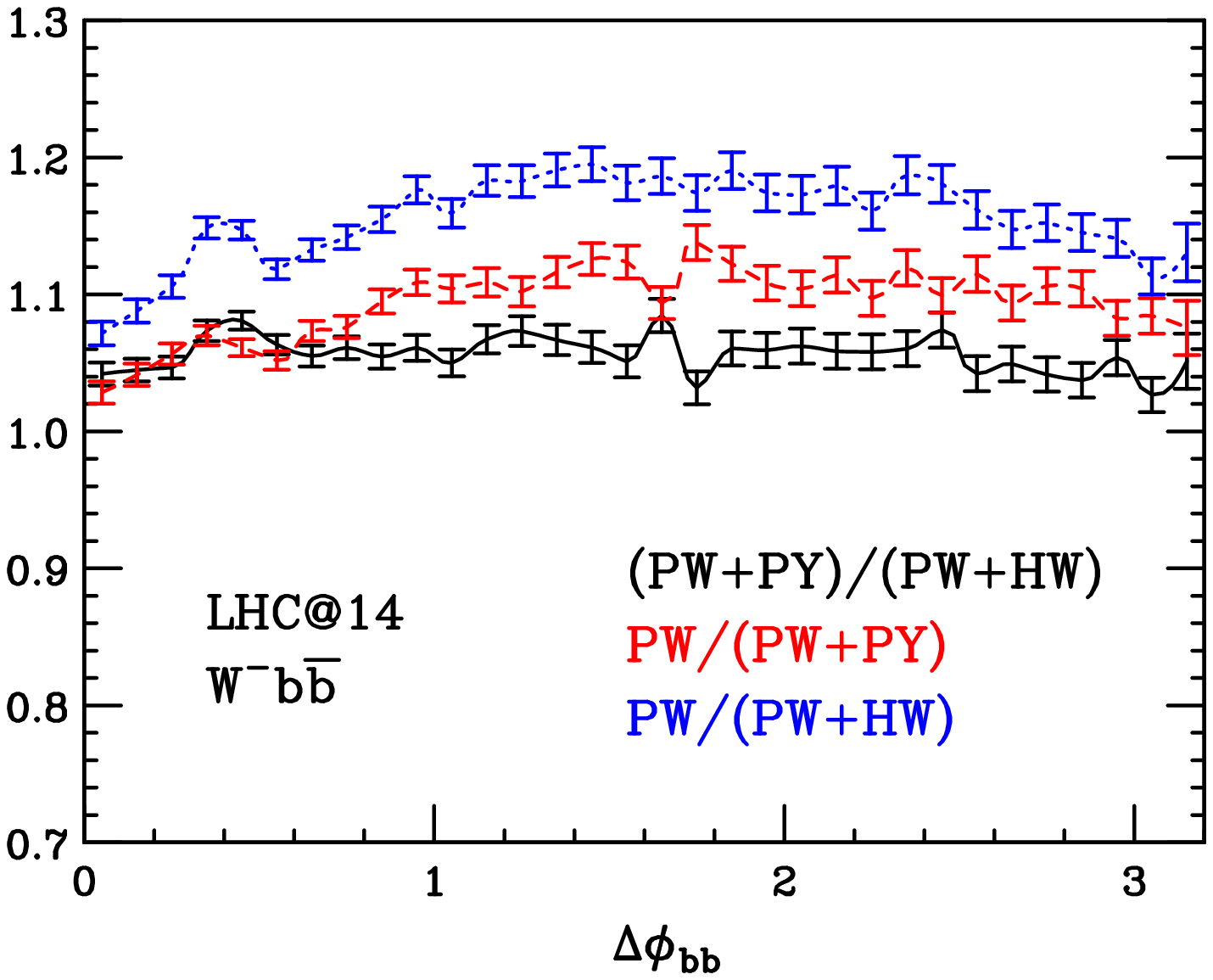,width=\wfig,height=\hfig}
\end{center}
\caption{\label{fig:LHC_PY_HW_ratio} Ratios of the differential cross
  sections for $\Wmbbdec$ production at the LHC:
  (\POWHEG{}+\PYTHIA{})/ (\POWHEG{}+\HERWIG{}) in solid black lines,
  \POWHEG{}/(\POWHEG{}+\PYTHIA{}) in dashed red lines and
  \POWHEG{}/(\POWHEG{}+\HERWIG{}) in dotted blue lines. Starting from the
  upper left corner and moving clockwise we show the ratio of the
  differential cross sections as function of the transverse momentum of the
  hardest $b$ jet, $p_T^b$ leading, the pseudorapidity of the second hardest
  $b$ jet, $\eta^b$ subleading, the azimuthal angular difference between the
  two $b$ jets, $\Delta\phi_{bb}$, and the rapidity of the hardest radiated
  non $b$ jet, $y^j$.}
\end{figure}

In fig.~\ref{fig:LHC_PY_HW_ratio}, we have plotted the ratios {\tt
  (PW+PY)/(PW+HW)} in solid black lines, {\tt PW/(PW+PY)} in dashed red lines
and {\tt PW/(PW+HW)} in dotted blue lines. The effects of the two different
showers, i.e.~the ratios {\tt (PW+PY)/(PW+HW)} are of the order of less than
10\% for most of the distributions considered, so that the differences
between the two showering algorithms is less pronounced at the LHC 
than at the Tevatron.  Again, the distribution that turns out to be more
sensitive to the showering procedure is the rapidity of the hardest jet, as
illustrated in the lower left-hand-side plot of the figure, where jets from
the \HERWIG{} shower tend to be more central in rapidity than jets from the
\PYTHIA{} shower.

\section{Conclusions}
\label{sec:conclusions}
In this article we have presented a next-to-leading order plus parton shower
simulation of the production of a $W$ boson in association with a massive
$b\bar{b}$ pair, based on the \POWHEG{} formalism, with the leptonic decay of
the $W$ boson taken into account using standard approximated techniques.  We
have assembled our generator with the aid of the \POWHEGBOX{}
toolkit~\cite{Alioli:2010xd}.  The NLO virtual corrections have been taken
from~\cite{FebresCordero:2006sj, Cordero:2008ce, Cordero:2009kv} and their
validity has been expanded in order to account for the case of an off-shell
$W$ boson production.
 
We have validated the code after taking care of activating the \POWHEGBOX{}
mechanism to protect form Born-zero configurations, and after separating out
from the part of the real-radiation contribution treated with Monte Carlo
techniques the region of hard gluons collinear to the final-state massive $b$
quarks, to be treated with standard NLO techniques. This was done to prevent
the enhancement of mass logarithmic terms by the \POWHEG{} factor
$\bar{B}/B$.

Finally, we have showered the hardest-emission results generated by \POWHEG{}
with two popular shower Monte Carlo programs: \PYTHIA{} and \HERWIG{}.
Looking at various kinematic distributions, we have found discrepancies of
the order of 10--20\% for the Tevatron and of less than 10\% for the LHC
between the two shower Monte Carlo programs.  Discrepancies larger than the
quoted values can be found in distributions involving the rapidity of the
hardest radiated jet.  These discrepancies can be considered as theoretical
errors associated with the two different showering algorithms.

The tool we provide will be very important for both Higgs-boson and beyond
the Standard Model searches at both the Tevatron and the LHC. Indeed $W+b$
jets is one of the main backgrounds to these searches and $Wb\bar{b}$
production is the main contribution.

\vspace{0.5cm}
\noindent
The code of our generator can be accessed in the \POWHEGBOX{} svn
repository\\
\centerline{\url{svn://powhegbox.mib.infn.it/trunk/POWHEG-BOX},}\\ with
username {\tt anonymous} and password {\tt anonymous}.

\acknowledgments 
We wish to thank P.~Nason and F.~Febres Cordero for useful discussions and we
acknowledge the use of the Turing cluster of the INFN group IV in the Theory
Division of the Physics Department at the University of Milano-Bicocca. We
thank E.~Re for carefully reading the paper. The work of L.R. is supported in
part by the U.S.~Departmet of Energy under grant DE-FG02-97IR41022.

\bibliography{paper}

\end{document}